\documentclass[oldversion]{aa} % double column
\usepackage{graphicx}
\usepackage{longtable}
\usepackage{txfonts}
\usepackage{rotating}
\usepackage{lscape}
\newcommand{\gsim}{\;\lower.6ex\hbox{$\sim$}\kern-7.75pt\raise.65ex\hbox{$>$}\;}
\newcommand{\lsim}{\;\lower.6ex\hbox{$\sim$}\kern-7.75pt\raise.65ex\hbox{$<$}\;}

\begin{document}
\title{Spectroscopic analysis of the two subgiant branches of the globular cluster NGC~1851
\thanks{Based on observations collected at ESO telescopes under programme 084.D-0470 }
 }

\author{
R.G. Gratton\inst{1},
S. Villanova\inst{2},
S. Lucatello\inst{1},
A. Sollima\inst{1},
D. Geisler\inst{2},
E. Carretta\inst{3},
S. Cassisi\inst{4},
\and
A. Bragaglia\inst{3}
}

\authorrunning{R.G. Gratton}
\titlerunning{Abundances in subgiants of NGC~1851}

\offprints{R.G. Gratton, raffaele.gratton@oapd.inaf.it}

\institute{
INAF-Osservatorio Astronomico di Padova, Vicolo dell'Osservatorio 5, I-35122
 Padova, Italy
\and
Departamento de Astronomia, Casilla 160, Universidad de Concepcion, Chile
\and
INAF-Osservatorio Astronomico di Bologna, Via Ranzani 1, I-40127
 Bologna, Italy
\and
INAF-Osservatorio Astronomico di Collurania, Via Maggini snc, 64100 Teramo, Italy
}

\date{}
\abstract{It has been found that globular clusters host multiple stellar
populations. In particular, in NGC~1851 the subgiant branch (SGB) can be divided
into two components and the distribution of stars along the horizontal 
branch (HB) is multimodal. Various authors have found that NGC~1851 
possibly has a spread in [Fe/H], but the relation between this spread and the
division in the SGB is unknown. We obtained blue (3950-4600~\AA) intermediate 
resolution ($R\sim 8,000$) spectra for 47 stars on the bright and 30 on the faint 
SGB of NGC~1851 (b-SGB and f-SGB, respectively). The spectra were analysed by 
comparing with synthetic spectra. The determination of the atmospheric parameters 
to extremely high internal accuracy allows small errors to be recovered when 
comparing different stars in the cluster, in spite of their faintness ($V\sim 19$). Abundances were 
obtained for Fe, C, Ca, Cr, Sr, and Ba. We found that the b-SGB is slightly more 
metal-poor than the f-SGB, with [Fe/H]=$-1.227\pm 0.009$\ and [Fe/H]=$-1.162\pm 0.012$, 
respectively. This implies that the f-SGB is only slightly older by $\sim 0.6$~Gyr 
than the b-SGB if the total CNO abundance is constant. There are more C-normal stars 
in the b-SGB than in the f-SGB. This is consistent with what is found for HB stars, 
if b-SGB are the progenitors of red HB stars, and f-SGB those of blue HB ones. As
previously found, the abundances of the n-capture elements Sr and Ba have a bimodal 
distribution, reflecting the separation between f-SGB (Sr and Ba-rich) and b-SGB stars 
(Sr and Ba-poor). In both groups, there is a clear correlation between [Sr/Fe] and 
[Ba/Fe], suggesting that there is a real spread in the abundances of n-capture elements. 
By looking at the distribution of SGB stars in the [C/H] vs. T$_{\rm eff}$ diagram 
and in the [Ba/H] vs. [Sr/H] diagram, not a one-to-one relation is found among these 
quantities. There is some correlation between C and Ba abundances, while the same correlation for Sr is much 
more dubious. We identified six C-rich stars, which have a moderate overabundance of 
Sr and Ba and rather low N abundances. This group of stars might be the progenitors of 
these on the anomalous RGB in the $(v, v-y)$\ diagram. These results are discussed
within different scenarios for the formation of NGC~1851. It is possible that the two 
populations originated in different regions of a inhomogeneous parent object. However, 
the striking similarity with M~22 calls for a similar evolution for these two clusters. 
Deriving reliable CNO abundances for the two sequences would be crucial.}

\maketitle

\section{Introduction}

Most, possibly all, globular clusters host multiple stellar populations (see reviews in
Gratton et al. 2004, 2012a). The most sensitive diagnostics of these populations is
the Na-O anticorrelation (Kraft 1994; Gratton et al. 2004), although very important 
information is also provided by the colour-magnitude diagram (see Piotto 2009; Gratton 
et al. 2010). In the majority of cases, available observations can be explained by two 
generations of stars: a primordial population, with the same chemical composition as 
field halo stars of similar metallicity; and a second generation, born from the ejecta 
of a fraction of the stars of the earlier one. The nature of the first generation of
stars whose ejecta are used to form the second generation is still debated: they may be
massive AGB stars undergoing hot bottom burning (see e.g. Ventura et al. 2001) or
rapidly rotating massive stars (see e.g. Decressin et al. 2007; etc.). In addition,
it is clear that the whole process is 
driven mainly by the total mass of the cluster (Carretta et al. 2010a), even though other 
parameters might have a role (location in the galaxy, orbit, etc.). 
However, there are a few cases that do not easily fit in 
this scenario and possibly require additional mechanisms to create multiple stellar 
populations. One of these anomalies is related to the splitting of the subgiant branch 
(SGB) which is observed in a few clusters, the most notable cases being NGC~1851 (Milone 
et al. 2008, 2009; Han et al. 2009) and M22 (=NGC~6656: Piotto 2009). This splitting might 
be attributed to either a quite large difference in age (of the order of 1 Gyr: Milone et al. 
2008) or CNO content (by at least a factor of two: Cassisi et al. 2008). Both of these 
features cannot be easily related to the usual second generation scenario, where the age 
difference between the first/primordial and the second generation is thought to be less 
than 100 Myr, and no significant alteration of the sum of CNO elements is foreseen (see 
Carretta et al. 2005; Villanova et al. 2010, Yong et al. 2011). This calls for additional 
mechanisms that are not yet understood.

Both NGC~1851 and M22 have been studied quite extensively in the past few years. High
dispersion spectra have been obtained for red giants (Yong \& Grundahl 2008; Villanova 
et al. 2010; Carretta et al. 2010c, 2011a; Marino et al. 2011; Roederer et al. 2011) and 
horizontal branch (HB) stars (Gratton et al. 2012b). Marino et al. (2012) and Lardo et 
al. (2012) also obtained intermediate and low resolution spectra of SGB stars in M22 and 
NGC~1851, respectively. In addition, Str\"omgren photometry was obtained for red 
giant branch (RGB) stars by Grundahl et al. (1999) and Lee et al. (2009), and recently 
discussed by Carretta et al. (2011a, 2011b) and Lardo et al. (2012). This quite impressive 
dataset showed that the two clusters (which differ in metallicity by some 0.5 dex) appear 
to share numerous observational features. In both cases, there is evidence of a spread in 
metal abundance that is larger in M22 ($\sim 0.2$ dex in [Fe/H]) but also present in NGC~1851 
(Lee et al 2009; Carretta et al. 2010b, 2010c, 2011a), although Villanova et al. (2010) 
found no evidence of an Fe spread among RGB stars from the split RGB. The neutron-capture 
elements that are mainly produced by the $s$-process are also found to have widely different 
star-to-star abundances that are correlated with both [Fe/H] and the abundances of light, 
proton-capture elements (Carretta et al. 2011a). Finally, a spread in the abundances of 
individual CNO elements has been found within both the bright and faint SGB (b-SGB and f-SGB, 
respectively), that is also correlated with the variation in heavier elements (Marino et al. 2012; 
Lardo et al. 2012).

NGC~1851 has additional peculiarities that have not yet been found in M22 where no 
bimodality in the HB exists, at least in the CMD, exists. The most obvious peculiarity
is the bimodal distribution of stars along the HB (Walker 1992): about 60\% of the stars
reside on the red horizontal branch (RHB), some 30\% being on the blue horizontal branch
(BHB), and only 10\% of the stars having intermediate colours that fall within the instability
strip. It is unclear whether the difference in the distribution of stars along the HB with
respect to M22 - where no bimodality has yet been discovered - might be explained simply 
by the different metallicity and age of the two clusters, or to some more significant 
difference in their formation scenarios. Gratton et al. (2012b) studied the chemical 
composition of about a hundred HB stars in NGC~1851. They found that both red and blue HB 
stars display separate Na-O anticorrelations, reinforcing the suggestion advanced by van 
den Bergh (1996), Catelan (1997), Carretta et al. (2010c, 2011a), and Bekki and Yong (2012) 
that NGC~1851 might possibly be the result of the merging of two clusters (another 
cluster which may be interesting in this context is Terzan 5, which also has a bimodal 
HB: see Ferraro et al. 2009 and Origlia et al. 2011). However, other observational
results remain unexplained. There is a group of RHB stars with high Ba (and Na)
abundances. These stars might possibly be related to the anomalous red sequence in the
Str\"omgren $(v, v-y)$\ diagram considered by Villanova et al. (2010) and Carretta et al. 
(2011a), which also exclusively consists of Ba-rich stars (Villanova et al. 2010), although 
Carretta et al. (2011a) found many other Ba-rich red giants in NGC~1851. A possible way of 
explaining the anomalous position of these stars in the $(v, v-y)$\ diagram is to assume that 
they have a larger CNO content (see Carretta et al. 2011c and Gratton et al. 2012b). The 
sum of the abundances of CNO elements of these stars in NGC~1851 is debated: Yong \& 
Grundahl (2008) and Yong et al. (2011; see Alves-Brito et al., 2012, for a similar claim 
for M~22) found indications of a large enhancement, while Villanova et al. (2010) instead found 
that they have a constant, low/normal CNO content. In all cases, this sequence contains
$\sim 10-15$\% of the stars, and can thus correspond directly to neither the f-SGB sequence 
(which consists of some 30-40\% of the SGB stars of NGC~1851: Milone et al. 2008, 2009) nor the 
BHB (which makes up a similar fraction of the HB stars: Milone et al. 2009). 

On the other hand, Han et al. (2009) found that the progeny of the f-SGB can be followed 
throughout the RGB in the $(U,U-I)$ diagram. A very similar result was obtained by 
Lardo et al. (2012) by using a particular combination of Str\"omgren colours ($(u+v)-(b+y)$). 
This combination, which compares the sum of the ultraviolet (UV) and violet magnitudes with 
sum of the blue and yellow ones, is actually conceptually
similar to the Han et al. broad band photometry. In both cases, the clear separation of 
stars in these sequences might be caused by a combination of different effective temperatures and 
absorption by atomic lines and molecular bands in the UV, so that its interpretation in 
terms of abundances is not simple: Han et al. attributed the difference to heavy elements, 
Lardo et al. to CNO, and Carretta et al. (2010b, 2011b) found that it is very 
well-correlated with O and Na abundances on the upper part of the RGB. Lardo et al. found 
that summing up these stars over the whole RGB, the fraction is $\sim 30$\%, as for 
f-SGB stars. Stars on the anomalous RGB in the $(v, v-y)$\ diagram are however only a 
subset of the "red" stars in the $(U,U-I)$ or $(u, (u-y)+(v-b))$\ diagrams, and should 
result from the evolution of only part of the f-SGB stars.

A detailed study of the chemical composition of stars on both the bright (b-SGB) and faint
(f-SGB) sequences of NGC~1851 might help us to understand the evolutionary history of this 
cluster, providing 
constraints on the ages and chemical compositions of its various populations. The SGB is 
possibly the part of the colour-magnitude diagram where subtle age differences may best 
be proven, but at present we do not even know the metallicity ranking of the two most 
important populations present in this cluster. Metallicity is a basic ingredient in age 
derivations, and even small abundance differences may have an impact in a case such as 
NGC~1851. Unfortunately, SGB stars are quite faint in such a distant cluster, so that high
resolution spectra would require a prohibitively long exposure time even on eight meter class
telescopes. Lardo et al. (2012) presented an analysis of low-resolution spectra for a total
of 64 stars near the SGB of NGC~1851: they found separate C-N anticorrelations for stars 
on the two branches, and evidence that on average the sums of C+N differ by a factor of 2.5 
between them. However, the spectra they used have a low resolution ($R\sim 1,000$) and 
they adopted a temperature scale that is apparently inconsistent with the evolutionary 
status of the stars observed. These facts cast doubts on their conclusions. 

In this paper, we present the results of the analysis of new intermediate-resolution 
($R\sim 8,000$) blue spectra of about eighty stars on the SGB of NGC~1851. These 
intermediate resolution spectra do not allow us to derive the abundances from weak lines, 
hence we are limited in the available diagnostics. However, we show that very useful 
information can be gathered once adequate care is taken in critical aspects of the analysis. 
The structure of this paper is as follows: the observational material is presented in 
Section 2; the analysis methods - which are quite unusual and were specially tailored for 
our case - are described in Section 3; our results are presented in Section 4. Discussion
and conclusions are drawn in Section 5.

\begin{center}
\begin{figure}
\includegraphics[width=8.8cm]{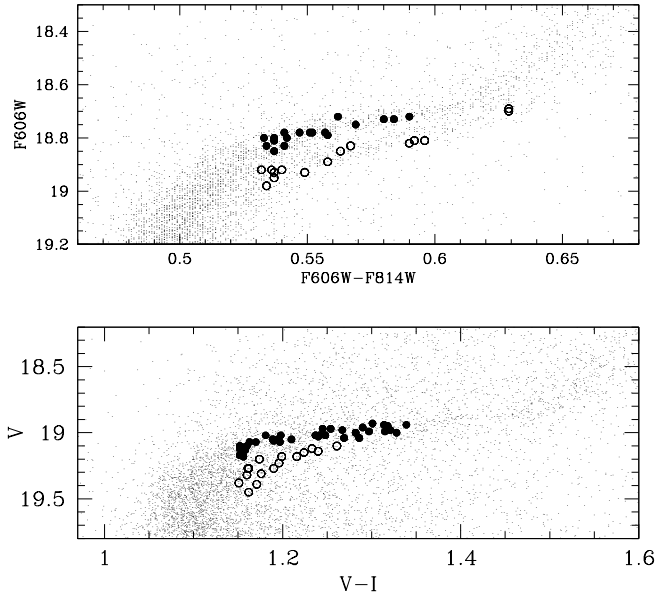}
\caption{Top: Advanced Camera for Surveys colour-magnitude diagram for the subgiant branch of NGC~1851 from Milone et
al. (2008). Bottom: The same, but using ground-based data.
Observed stars on the b-SGB (filled circles) and f-SGB (open circles) are
shown.}
\label{f:fig0}
\end{figure}
\end{center}

\begin{center}
\begin{figure}
\includegraphics[width=8.8cm]{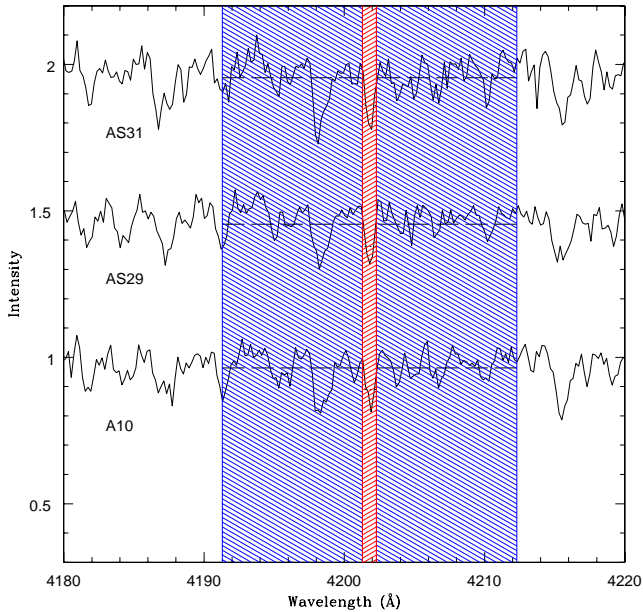}
\caption{Portion of the spectra of three SGB stars of NGC~1851, including the strong Fe line
at 4202~\AA\ (AS31, AS29, and A10). Spectra of the first two stars have been shifted vertically
for clarity. This is one of the fifteen strong Fe lines used in our Fe abundance derivation.
The areas hatched in red and blue mark the regions used to evaluate the line index appropriate 
for this line as described in the text (red is the in-line region, blue is the reference one). 
The average flux within each of these regions are shown as horizontal dashed lines. }
\label{f:fig3b}
\end{figure}
\end{center}

\section{Observations and data reduction}

Target stars on both the two SGBs were selected from the Advanced Camera
for Survey (ACS) of the Hubble Space Telescope (HST) (Milone et al. 2008) 
and ground-based $VI$\ photometry performed by Stetson, as described in Milone et al. (2009). In 
the first case, the high-precision photometry allowed a straightforward assignment 
of each star to its respective SGB, in the second some degree of uncertainty is 
present because the two SGBs are not so clearly separated, especially in the 
reddest part ($V-I>1.3$). Observed objects are shown in Fig.~\ref{f:fig0} as filled
circles (b-SGB) and open circles (f-SGB), respectively.

The spectra were acquired with the GIRAFFE fibre-fed spectrograph at the Very
Large Telescope (VLT) (Pasquini et 
al. 2004) under the ESO programme 084.D-0470 (PI Villanova), using the LR02 grating. They 
cover approximately the wavelength range 3950-4560~\AA\ at a resolution of about 
0.5~\AA\ (FWHM). In this spectral range, there are many strong lines of H, Fe, Ca, Sr, 
Ba, and CH, as well as many other lines that were not used in the present analysis. 
Targets are quite faint (V$\sim$19), so for each we obtained 20$\times$45 min
spectra in order to reach the required signal-to-noise (S/N) ($\sim$50). The spectra were reduced by
the ESO personnel using the ESO FLAMES GIRAFFE pipeline version 2.8.7, and then 
rectified to an approximate continuum value. This was done in a homogeneous way for 
all spectra, and turned out to be very useful for our analysis. However, we should take 
into account the possibility that some systematic difference is induced by this rectification 
depending on
for instance S/N of the spectra or temperature of the stars. In all cases, this 
"rectification" of the spectra does not have the same meaning as tracing a continuum in high 
dispersion spectra, because at the resolution of the spectra no pixels remain unaffected 
by absorption of some spectral lines. This was considered in our analysis. 

We measured radial velocities using the {\it fxcor} package in IRAF\footnote{IRAF is
distributed by the National Optical Astronomical Observatory, which are operated by 
the Association of Universities for Research in Astronomy, under contract with the 
National Science Foundation}, adopting a synthetic spectrum as a template. All stars 
turned out to have a velocity compatible with that of the cluster confirming their 
membership. We obtained an average radial velocity of $318.2\pm 0.5$~km s$^{-1}$\, with 
a root mean square (r.m.s.) scatter of 4.3~km~s$^{-1}$; very similar results were obtained by considering
separately b-SGB and f-SGB stars. For comparison, Scarpa et al. (2011) found an average 
velocity of $320.0\pm 0.4$~km s$^{-1}$\ with an r.m.s. of 4.9 km s$^{-1}$\ for 184 SGB 
stars, and that the cluster is slowly rotating (see also Bellazzini et al. 2012). Similar 
values were obtained by Carretta et al. (2011b) for stars on the RGB ($320.3\pm 
0.4$~km s$^{-1}$, r.m.s. of 3.7~km~s$^{-1}$), and by Gratton et al. 
(2012b) for stars on the RHB ($319.7\pm 0.5$~km s$^{-1}$, r.m.s. of 3.7~km~s$^{-1}$), BHB 
($321.6\pm 0.7$~km s$^{-1}$, r.m.s. of 4.1~km~s$^{-1}$), and lower RGB ($320.3\pm 
1.0$~km s$^{-1}$, r.m.s. of 3.6~km~s$^{-1}$). The offsets between these different sets 
of radial velocity data are nominally larger than the statistical errors. In principle, we 
might expect to find small offsets between the average radial velocities for stars in different 
evolutionary phases that are produced by convective motions and gravitational redshifts; however, 
in the present study they are most likely due to systematic differences among results obtained 
using different gratings and templates.
Before any additional step could be made, the spectra were then reduced to zero radial velocity. 

A small portion of each of the spectra of three stars is shown in Figure~\ref{f:fig3b}.

\begin{center}
\begin{figure}
\includegraphics[width=8.8cm]{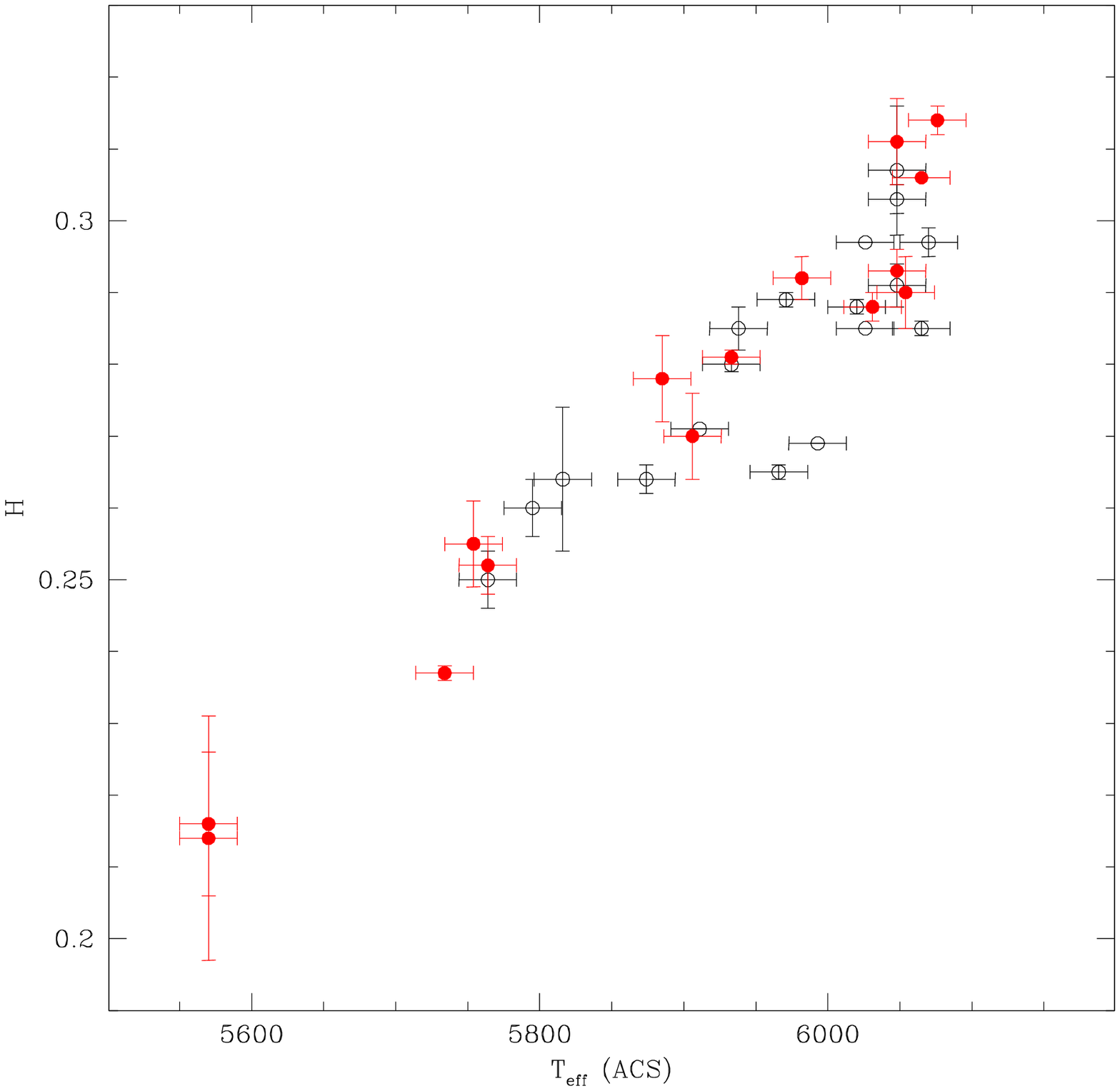}
\caption{Comparison of temperatures derived using the ACS F606W-F814W colour and the 
parameter H describing the strengths of H$\gamma$\ and H$\delta$.
Filled red symbols are for f-SGB stars, and empty black symbols represent b-SGB ones.}
\label{f:fig1}
\end{figure}
\end{center}

\begin{center}
\begin{figure}
\includegraphics[width=8.8cm]{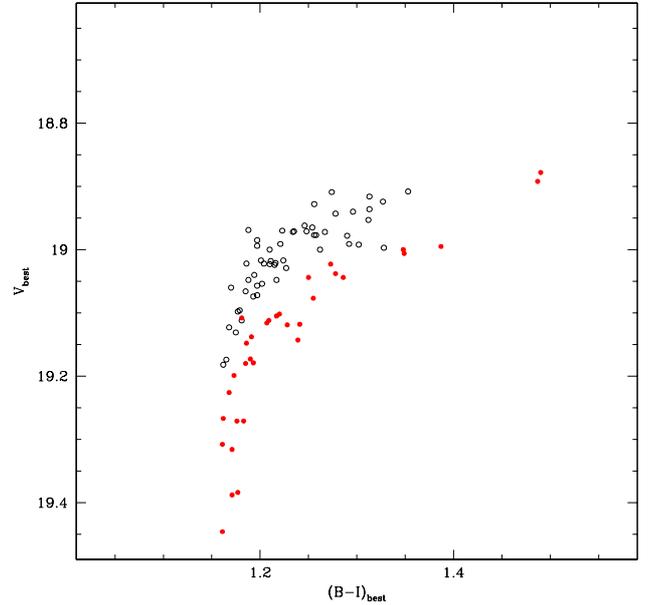}
\caption{Pseudo colour-magnitude diagram obtained using our pseudo-colour 
$(B-I)_{\rm best}$. Symbols are as in Figure~\ref{f:fig1}. }
\label{f:fig2}
\end{figure}
\end{center}

\begin{center}
\begin{figure}
\includegraphics[width=8.8cm]{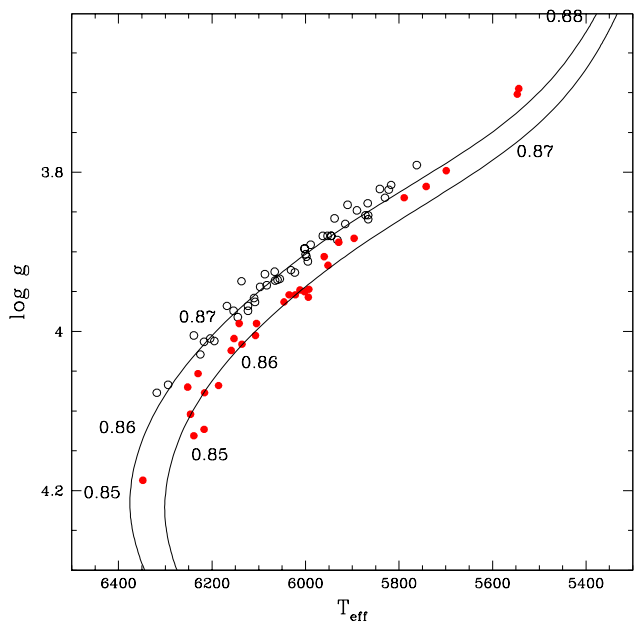}
\caption{Distribution of the programme stars in the effective temperature-surface gravity plane. 
Symbols are as in Figure~\ref{f:fig1}. We overimpose two 
$\alpha$-enhanced isochrones from the BASTI database (Pietrinferni et al. 2006), which were computed for [Fe/H] values 
appropriate to the two sequences ([Fe/H]=-1.22 and -1.17) and ages of 10.1 and 10.7 Gyr 
for b-SGB and f-SGB, respectively. Stellar masses along these isochrones are also given.  }
\label{f:fig3}
\end{figure}
\end{center}

\section{Analysis}

\begin{table*}[htb]
\centering
\caption[]{Atmospheric parameters and abundances in the program stars - Bright SGB}
\begin{tabular}{lcccccccccc}
\hline
ID  &$T_{\rm eff}$&	$\log{g}$ &	$v_t$ &	$[$Fe/H$]$ & $[$C/Fe$]$ & $[$Ca/Fe$]$& $[$Cr/Fe$]$ & $[$Sr/Fe$]$ & $[$Ba/Fe$]$ & $\Delta u-y$\\
    & (K)         &           &(km s$^{-1}$) \\
\hline
A10 	&$6002\pm 21$ &	3.895&	0.97&$-1.23\pm 0.04$&$-0.08\pm 0.03$&	0.30& -0.36 &$~~0.14\pm 0.10$&	~~0.81 & -0.016 \\
A12 	&$5989\pm 21$ &	3.890&	0.97&$-1.22\pm 0.04$&$~~0.11\pm 0.01$&	0.49& -0.16 &$-0.10\pm 0.20$&	~~1.02 & ~~0.108\\
A13 	&$5915\pm 21$ &	3.864&	0.98&$-1.42\pm 0.03$&$~~0.25\pm 0.01$&	0.43&~~0.02 &$~~0.15\pm 0.06$&	~~0.95 & ~~0.054\\
A16 	&$5953\pm 21$ &	3.879&	0.97&$-1.20\pm 0.03$&$-0.12\pm 0.02$&	0.41& -0.36 &$~~0.27\pm 0.05$&	~~1.10 & ~~0.325\\
A17 	&$6000\pm 21$ &	3.902&	0.96&$-1.13\pm 0.03$&$-0.03\pm 0.02$&	0.31& -0.02 &$-0.09\pm 0.17$&	~~0.56  & ~~0.133\\
A21 	&$5762\pm 21$ &	3.790&	1.00&$-1.14\pm 0.03$&$~~0.01\pm 0.03$&	0.45& -0.04 &$~~0.11\pm 0.02$&	~~0.88 & ~~0.001\\
A23 	&$6087\pm 21$ &	3.927&	0.96&$-1.26\pm 0.04$&$~~0.26\pm 0.02$&	0.31& -0.10 &$-0.01\pm 0.10$&	~~0.94 & ~~0.062\\
A24 	&$5817\pm 21$ &	3.815&	0.99&$-1.18\pm 0.03$&$~~0.08\pm 0.03$&	0.36& -0.41 &$~~0.01\pm 0.06$&	~~0.69 & -0.015 \\
A25 	&$5945\pm 21$ &	3.879&	0.97&$-1.22\pm 0.02$&$-0.01\pm 0.02$&	0.32& -0.18 &$~~0.23\pm 0.00$&	~~1.04 & ~~0.002 \\
A27 	&$5872\pm 21$ &	3.853&	0.98&$-1.11\pm 0.03$&$-0.05\pm 0.02$&	0.27& -0.11 &$~~0.18\pm 0.03$&	~~0.58  & ~~0.004\\
A29 	&$5998\pm 21$ &	3.905&	0.96&$-1.26\pm 0.04$&$~~0.10\pm 0.02$&	0.37& -0.22 &$-0.03\pm 0.13$&	~~0.83 & ~~0.099\\
A30 	&$5995\pm 21$ &	3.911&	0.96&$-1.20\pm 0.03$&$~~0.08\pm 0.02$&	0.37& -0.08 &$~~0.18\pm 0.07$&	~~0.74 & ~~0.083\\
A34 	&$5867\pm 21$ &	3.838&	0.98&$-1.20\pm 0.03$&$~~0.09\pm 0.04$&	0.42& -0.24 &$~~0.14\pm 0.03$&	~~0.52  & ~~0.057\\
A35 	&$6060\pm 21$ &	3.934&	0.95&$-1.25\pm 0.03$&$-0.05\pm 0.01$&	0.34& -0.03 &$~~0.17\pm 0.17$&	~~0.93 & ~~0.108\\
A03	 	&$5841\pm 21$ &	3.820&	0.99&$-1.16\pm 0.03$&$-0.13\pm 0.01$&	0.44& -0.26 &$~~0.16\pm 0.03$&	~~0.83 & -0.034 \\
A04	 	&$6110\pm 21$ &	3.957&	0.95&$-1.21\pm 0.05$&$-0.15\pm 0.05$&	0.34& -0.10 &$~~0.21\pm 0.01$&	~~0.99 & ~~0.120\\
A06	 	&$5910\pm 21$ &	3.840&	0.98&$-1.31\pm 0.04$&$~~0.14\pm 0.02$&	0.37& -0.31 &$-0.04\pm 0.09$&	~~0.89 & -0.039 \\
A09	 	&$6066\pm 21$ &	3.924&	0.96&$-1.30\pm 0.03$&$~~0.23\pm 0.00$&	0.38& -0.19 &$~~0.28\pm 0.07$&	~~0.86 & -0.040 \\
AS12	&$5830\pm 23$ &	3.831&	0.99&$-1.27\pm 0.03$&$-0.19\pm 0.00$&	0.37& -0.32 &$~~0.02\pm 0.12$&	~~0.70 & -0.055 \\
AS14	&$5866\pm 23$ &	3.858&	0.98&$-1.21\pm 0.03$&$~~0.11\pm 0.03$&	0.38& -0.38 &$~~0.08\pm 0.10$&	~~0.93 & -0.017 \\
AS15	&$6097\pm 23$ &	3.943&	0.95&$-1.15\pm 0.03$&$~~0.17\pm 0.03$&	0.31&~~0.14 &$~~0.28\pm 0.02$&	~~1.03 &        \\
AS18	&$6031\pm 23$ &	3.922&	0.96&$-1.24\pm 0.03$&$-0.16\pm 0.04$&	0.39& -0.29 &$~~0.17\pm 0.01$&	~~0.81 & ~~0.040\\
AS22	&$6002\pm 23$ & 3.895&	0.97&$-1.25\pm 0.03$&$~~0.22\pm 0.02$&	0.42&~~0.05 &$~~0.19\pm 0.09$&	~~0.88 &        \\
AS24	&$5866\pm 23$ &	3.853&	0.98&$-1.21\pm 0.04$&$-0.01\pm 0.00$&	0.46& -0.07 &$~~0.06\pm 0.03$&	~~1.04 & -0.012 \\
AS25	&$5945\pm 23$ &	3.879&	0.97&$-1.21\pm 0.04$&$~~0.09\pm 0.03$&	0.44& -0.14 &$~~0.15\pm 0.08$&	~~0.93 & -0.151 \\
AS28	&$6168\pm 23$ &	3.967&	0.94&$-1.30\pm 0.04$&$~~0.37\pm 0.01$&	0.58&~~0.35 &$~~0.32\pm 0.16$&	~~0.83 & ~~0.037\\
AS29	&$6204\pm 23$ &	4.008&	0.93&$-1.32\pm 0.05$&$~~0.31\pm 0.03$&	0.43&~~0.21 &$~~0.52\pm 0.04$&	~~1.01 & ~~0.031\\
AS02	&$6318\pm 23$ &	4.076&	0.91&$-1.11\pm 0.04$&$~~0.07\pm 0.06$&	0.21& -0.02 &$-0.02\pm 0.23$&	~~0.81 &        \\
AS31	&$5822\pm 23$ &	3.821&	0.99&$-1.21\pm 0.04$&$~~0.14\pm 0.03$&	0.37& -0.19 &$~~0.07\pm 0.06$&	~~0.80 &        \\
AS36	&$6217\pm 23$ &	4.012&	0.93&$-1.33\pm 0.06$&$~~0.33\pm 0.05$&	0.59&~~0.51 &$~~0.50\pm 0.08$&	~~1.20 &        \\
AS37	&$6065\pm 23$ &	3.935&	0.95&$-1.17\pm 0.03$&$-0.04\pm 0.03$&	0.30& -0.33 &$~~0.14\pm 0.03$&	~~0.85 & -0.016 \\
AS38	&$5938\pm 23$ &	3.857&	0.98&$-1.22\pm 0.03$&$~~0.14\pm 0.01$&	0.24& -0.02 &$~~0.12\pm 0.09$&	~~0.82 &        \\
AS39	&$5932\pm 23$ &	3.884&	0.97&$-1.30\pm 0.04$&$~~0.23\pm 0.03$&	0.48& -0.25 &$~~0.47\pm 0.02$&	~~1.03 & -0.072 \\
AS40	&$6294\pm 23$ &	4.066&	0.91&$-1.18\pm 0.05$&$~~0.21\pm 0.02$&	0.60&~~0.24 &$~~0.42\pm 0.07$&	~~1.19 & ~~0.004\\
AS42	&$6108\pm 23$ &	3.962&	0.94&$-1.30\pm 0.06$&$~~0.31\pm 0.04$&	0.27&~~0.24 &$~~0.34\pm 0.05$&	~~1.26 & -0.013 \\
AS43	&$6225\pm 23$ &	4.028&	0.92&$-1.24\pm 0.04$&$~~0.32\pm 0.03$&	0.52& -0.16 &$~~0.04\pm 0.00$&	~~0.83 & ~~0.055\\
AS44	&$5890\pm 23$ &	3.847&	0.98&$-1.23\pm 0.03$&$~~0.03\pm 0.02$&	0.47& -0.14 &$~~0.23\pm 0.02$&	~~0.92 & -0.028 \\
AS45	&$6154\pm 23$ &	3.973&	0.94&$-1.17\pm 0.05$&$~~0.21\pm 0.00$&	0.41&~~0.10 &$~~0.11\pm 0.08$&	~~0.84 & -0.048 \\
AS46	&$6195\pm 23$ &	4.011&	0.93&$-1.19\pm 0.05$&$~~0.25\pm 0.01$&	0.33& -0.19 &$~~0.03\pm 0.06$&	~~0.79 & -0.007 \\
AS48	&$6145\pm 23$ &	3.981&	0.94&$-1.22\pm 0.04$&$~~0.21\pm 0.03$&	0.31& -0.19 &$~~0.11\pm 0.02$&	~~0.79 & -0.045 \\
AS04	&$6123\pm 23$ &	3.973&	0.94&$-1.25\pm 0.05$&$~~0.15\pm 0.03$&	0.33& -0.12 &$~~0.35\pm 0.11$&	~~1.07 & -0.015 \\
AS50	&$6239\pm 23$ &	4.004&	0.93&$-1.20\pm 0.05$&$~~0.31\pm 0.02$&	0.61& -0.21 &$~~0.12\pm 0.07$&	~~1.09 &        \\
AS54	&$6083\pm 23$ &	3.941&	0.95&$-1.16\pm 0.04$&$~~0.26\pm 0.02$&	0.46&~~0.09 &$~~0.46\pm 0.05$&	~~1.09 &        \\
AS56	&$6123\pm 23$ &	3.967&	0.94&$-1.28\pm 0.04$&$~~0.06\pm 0.08$&	0.43&~~0.12 &$~~0.34\pm 0.03$&	~~0.70 &        \\
AS05	&$6023\pm 23$ &	3.925&	0.96&$-1.26\pm 0.06$&$-0.09\pm 0.04$&	0.35& -0.05 &$~~0.16\pm 0.12$&	~~0.66 &        \\
AS08	&$6055\pm 23$ &	3.933&	0.95&$-1.20\pm 0.04$&$-0.17\pm 0.09$&	0.23&~~0.08 &$~~0.12\pm 0.03$&	~~0.47 &        \\
AS09	&$5963\pm 23$ &	3.879&	0.97&$-1.27\pm 0.04$&$~~0.23\pm 0.03$&	0.37& -0.28 &$~~0.17\pm 0.04$&	~~0.77 &        \\
\hline 
\end{tabular}
\label{t:tab1a}
\end{table*}

\begin{table*}[htb]
\centering
\caption[]{Atmospheric parameters and abundances in the program stars - Faint SGB}
\begin{tabular}{lcccccccccc}
\hline
ID  &$T_{\rm eff}$&	$\log{g}$ &	$v_t$ &	$[$Fe/H$]$ & $[$C/Fe$]$ & $[$Ca/Fe$]$ &	$[$Cr/Fe$]$ &	$[$Sr/Fe$]$ & $[$Ba/Fe$]$ & $\Delta u-y$\\
    & (K)         &           &(km s$^{-1}$) \\
\hline
B10		&$5952\pm 21$ &	3.916&	0.96&$-1.24\pm 0.04$&$~~0.02\pm 0.05$&	0.40&~~0.06 &$~~0.41\pm 0.03$&	~~0.93 & ~~0.098\\
B11		&$6022\pm 21$ &	3.953&	0.95&$-1.23\pm 0.03$&$~~0.05\pm 0.02$&	0.50& -0.02 &$~~0.40\pm 0.02$&	~~1.03 & ~~0.066\\
B12		&$6046\pm 21$ &	3.962&	0.94&$-1.19\pm 0.03$&$~~0.20\pm 0.02$&	0.43&~~0.05 &$~~0.37\pm 0.03$&	~~1.20 & -0.048 \\
B13		&$5547\pm 21$ &	3.701&	1.03&$-1.04\pm 0.03$&$-0.37\pm 0.03$&	0.40& -0.06 &	             &         & ~~0.329\\
B15		&$6107\pm 21$ &	4.004&	0.93&$-1.13\pm 0.04$&$-0.19\pm 0.06$&	0.35& -0.02 &$~~0.38\pm 0.01$&	~~0.85 & ~~0.288\\
B16		&$6002\pm 21$ &	3.949&	0.95&$-1.13\pm 0.03$&$-0.20\pm 0.05$&	0.43& -0.17 &$~~0.40\pm 0.17$&	~~0.93 & ~~0.108\\
B18		&$5789\pm 21$ &	3.831&	0.99&$-1.15\pm 0.04$&$~~0.00\pm 0.04$&	0.48& -0.03 &$~~0.33\pm 0.10$&	~~1.06 & ~~0.044\\
B01		&$5544\pm 21$ &	3.694&	1.03&$-1.12\pm 0.03$&$-0.34\pm 0.04$&	0.59& -0.23 &	             &         & -0.004 \\
B02		&$6142\pm 21$ &	3.989&	0.94&$-1.25\pm 0.03$&$~0.03\pm 0.04$&	0.25& -0.07 &$~~0.17\pm 0.11$&	~~1.04 & ~~0.105\\
B03		&$5929\pm 21$ &	3.887&	0.97&$-1.10\pm 0.03$&$-0.15\pm 0.03$&	0.40& -0.05 &$~~0.33\pm 0.01$&	~~1.04 & ~~0.126\\
B04		&$5742\pm 21$ &	3.817&	0.99&$-1.13\pm 0.03$&$-0.24\pm 0.02$&	0.36& -0.15 &$~~0.30\pm 0.03$&	~~1.11 & ~~0.343\\
B05		&$6012\pm 21$ &	3.947&	0.95&$-1.22\pm 0.03$&$-0.11\pm 0.02$&	0.44& -0.18 &$~~0.45\pm 0.00$&	~~1.11 & ~~0.102\\
B06		&$6105\pm 21$ &	3.989&	0.94&$-1.09\pm 0.03$&$-0.11\pm 0.02$&	0.34&~~0.01 &$~~0.31\pm 0.01$&	~~1.02 & ~~0.168\\
B07		&$5896\pm 21$ &	3.882&	0.97&$-1.20\pm 0.04$&$-0.01\pm 0.02$&	0.40& -0.18 &$~~0.42\pm 0.05$&	~~1.12 & ~~0.139\\
B09		&$5699\pm 21$ &	3.797&	1.00&$-1.22\pm 0.03$&$-0.14\pm 0.00$&	0.39& -0.41 &$~~0.37\pm 0.24$&	~~1.15 & ~~0.069\\
BS10	&$6186\pm 23$ &	4.067&	0.91&$-1.12\pm 0.05$&$~~0.16\pm 0.03$&	0.17& -0.09 &$~~0.21\pm 0.00$&	~~0.57  & -0.003 \\
BS11	&$6136\pm 23$ &	4.015&	0.93&$-1.19\pm 0.04$&$-0.03\pm 0.05$&	0.37& -0.05 &$~~0.37\pm 0.05$&	~~1.11 & ~~0.003\\
BS12	&$6230\pm 23$ &	4.052&	0.92&$-1.08\pm 0.05$&$~~0.23\pm 0.01$&	0.40&~~0.03 &$~~0.26\pm 0.08$&	~~1.16 &        \\
BS14	&$6246\pm 23$ &	4.103&	0.90&$-1.16\pm 0.03$&$~~0.03\pm 0.03$&	0.49&~~0.10 &$~~0.46\pm 0.03$&	~~1.26 & ~~0.060\\
BS16	&$5960\pm 23$ &	3.905&	0.96&$-1.22\pm 0.05$&$~~0.44\pm 0.03$&	0.37&~~0.07 &$~~0.43\pm 0.01$&	~~1.27 & ~~0.128\\
BS17	&$6159\pm 23$ &	4.023&	0.92&$-1.22\pm 0.04$&$~~0.15\pm 0.06$&	0.43&~~0.04 &$~~0.53\pm 0.00$&	~~1.28 & ~~0.185\\
BS18	&$6153\pm 23$ &	4.008&	0.93&$-1.13\pm 0.04$&$~~0.07\pm 0.01$&	0.46&~~0.13 &$~~0.24\pm 0.52$&	~~1.09 &        \\
BS19	&$6348\pm 23$ &	4.186&	0.87&$-1.13\pm 0.05$&$~~0.36\pm 0.04$&	0.56&~~0.40 &$~~0.48\pm 0.04$&	~~1.35 &        \\
BS20	&$6252\pm 23$ &	4.069&	0.91&$-1.13\pm 0.05$&$~~0.22\pm 0.07$&	0.54&~~0.19 &$~~0.47\pm 0.00$&	~~1.25 &        \\
BS21	&$5993\pm 23$ &	3.946&	0.95&$-1.21\pm 0.03$&$-0.18\pm 0.06$&	0.39& -0.07 &$~~0.37\pm 0.04$&	~~1.16 &        \\
BS03	&$6217\pm 23$ &	4.122&	0.89&$-1.27\pm 0.03$&$~~0.14\pm 0.02$&	0.41&~~0.05 &$~~0.33\pm 0.05$&	~~1.11 & ~~0.021\\
BS04	&$6216\pm 23$ &	4.076&	0.91&$-1.22\pm 0.07$&$~~0.12\pm 0.03$&	0.46& -0.36 &$~~0.45\pm 0.05$&	~~1.30 &        \\
BS06	&$6035\pm 23$ &	3.953&	0.95&$-1.05\pm 0.03$&$-0.19\pm 0.05$&	0.36&~~0.10 &$~~0.32\pm 0.15$&	~~0.60 & ~~0.068\\
BS08	&$5994\pm 23$ &	3.956&	0.95&$-1.29\pm 0.05$&$~~0.39\pm 0.04$&	0.34&~~0.12 &$~~0.55\pm 0.29$&	~~1.13 & ~~0.013\\
BS09	&$6239\pm 23$ &	4.130&	0.89&$-1.00\pm 0.05$&$~~0.16\pm 0.07$&	0.26& -0.31 &$~~0.00\pm 0.36$&	~~0.90 & ~~0.022\\
\hline 
\end{tabular}
\label{t:tab1b}
\end{table*}
 
\subsection{Determination of the atmospheric parameters}

The main purpose of our analysis was to study the internal spread in the abundances 
within NGC~1851 and to look for systematic differences between the b- and f-SGB 
stars. We emphasize that we made no attempt to derive accurate absolute values for the average 
abundances within the cluster. Our goal requires the determination of atmospheric 
parameters with small errors when comparing different stars in similar evolutionary 
phases, while systematic errors affecting all programme stars in a similar way 
are less of a concern. However, we found that in spite of the large uncertainties in the 
abundances derived from blue spectra of intermediate resolution, the average 
abundances we obtained agree well with those found for stars in other evolutionary 
phases in this cluster and allow a consistent fit to the colour-magnitude diagram
(CMD). We are aware that this might be the result of our compensation for the errors, but
we still deem this as a satisfactory result.

The first step in our analysis was the derivation of effective temperatures; for 
practical reasons, they were ultimately based on the scale used in the BASTI isochrones 
(Pietrinferni et al. 2006)\footnote{http://albione.oa-teramo.inaf.it/}. 
We used a reddening of E(B-V)=0.02 as listed by Harris (1996) and a preliminary 
metallicity of [Fe/H]=-1.10, which is slightly higher than the final derived
value but this small difference only affects the absolute, not relative
values). Very accurate photometry in the F606W and F814W bands 
from ACS at HST is available from Milone et al. (2008, 2009). This photometry, however, 
only covers the central portion of the cluster, including a total of 33 target stars. 
For them, we obtained temperatures from this colour using the same calibration used 
by the BASTI database. Internal errors were derived by multiplying the median value of
the photometric error (0.0086~mag in the magnitude range of interest) by the 
sensitivity of the temperature to colour ($\sim 5000$~K/mag): the result is $\pm 43$~K.

For the remaining 53 stars, only ground-based photometry is available. While of good 
quality, this could not be compared with the HST photometry. Best results are obtained 
using $B-I$. A more relevant drawback for us, however, is that this photometry alone 
does not allow us to make an accurate distinction between along the f-SGB and b-SGB, 
with several ambiguous cases. Fortunately, our spectra contain several hydrogen lines,
particularly both H$\gamma$\ and H$\delta$, that can be measured accurately. We found 
that quite accurate temperatures can be obtained by combining these two lines. We 
measured the flux in bands 8~\AA\ wide centred on both lines, that had been normalized 
to the reference continuum value obtained as described in Section 3.2. 
We averaged these quantities, defining a parameter that we called $H$. An estimate of 
the typical error in this quantity was obtained by averaging quadratically the 
internal errors; these were obtained from the difference between the values we
obtained for H$\gamma$\ and H$\delta$. Since H$\gamma$\ on average yields a slightly larger 
value than H$\delta$\ (0.304 compared to 0.286), the last values were multiplied by this
ratio (i.e. 0.304/0.286) before estimating this difference. The average quadratic difference
is 0.010; if we attribute identical errors to the two quantities, the error in the
average is half this value, that is $\pm$0.005. We adopt this as our estimate of the internal
error in $H$. We found that $H$ is closely correlated with temperatures from F606W-F814W data
for the 33 stars with
available HST colours (see Fig~\ref{f:fig1}). We exploited this good correlation to 
calibrate the H parameter in terms of temperatures. Since the sensitivity is 5000 
K/unit change in $H$, the internal errors estimated above yield an error in temperature 
from $H$\ of $\pm 25$~K. The scatter in the values of H around the calibration line with 
temperatures from ACS photometry yields a mean quadratic difference of $\pm 49$~K, which 
is very close to the quadratic sum of the errors in the temperature derived from both the ACS 
photometry and $H$. This closely agree with our two estimates of the internal errors.

Our effective temperatures are the weighted averages of the values we get from the 
F606W-F814W ACS colour, the $H-$index, and $(B-I)$. These last temperatures were obtained 
from the ground-based $B-I$\ photometry described in Sect.~2, calibrated 
by using a best-fit linear relation with those provided by the remaining indices
\footnote{Temperatures from $(B-I)$\ obtained in this way differ from those
directly obtained from the calibration used in BASTI. The difference between these
two estimates of temperatures is well reproduced by a linear relation with
slope 1.246 and constant term -1320~K. In practice, the temperatures we adopted
are larger than those directly obtained from $(B-I)$\ using BASTI calibration by
168~K for the warmest stars, and are cooler by 15~K for the coolest stars of our sample.
While this difference has quite a large impact on trends of abundances with
effective temperatures, it does not affect relative abundances from b-SGB and f-SGB
stars, which have very similar average colours of $B-I=1.193$ and 1.195 for b-SGB and f-SGB,
respectively. We prefer our approach because it reproduces more closely the shape of the
SGB in the colour-magnitude diagram.}; 
the r.m.s. scatter around the best-fit line (77 K) indicates that these temperatures 
have errors of $\pm 65$~K, once the uncertainties in temperatures from other indices
are subtracted quadratically. The effective temperatures for the programme stars are listed 
in Tables~\ref{t:tab1a} and ~\ref{t:tab1b}, along with their internal errors. The systematic 
errors are likely much larger, but as mentioned above they are not of much relevance to our analysis.

For the sake of visualization, we plotted in Fig.~\ref{f:fig2} a colour-magnitude 
diagram obtained by transforming both F606W-F814W and H indices to the $(B-I)$\ 
system: we referred to our values of $(B-I)_{\rm best}$\ as these pseudo-colours. The b-SGB and the f-SGB 
sequences are well-defined. 

Surface gravities were obtained using the location of stars in the CMD. We 
assumed masses of 0.872~M$_\odot$\ for the b-SGB stars and 0.862~M$_\odot$\ for the f-SGB. 
These values were obtained after some iteration, by considering the effects of metal abundance 
and age when fitting the stars in the effective temperature-gravity plane with 
$\alpha$-enhanced ([$\alpha$/Fe]=+0.4) from the BASTI database. This procedure initiated 
from the apparent magnitudes that were corrected for the bolometric correction by 
Alonso et al. (1996), and then transformed into absolute magnitudes by using the distance 
modulus of Harris (1996). Fig.~\ref{f:fig3} shows the distribution of the stars in the effective 
temperature-surface gravity plane. The surface gravities likely have extremely small internal 
errors (typical values are $\pm 0.02$~dex) that are mainly due to the errors in the effective 
temperatures. We note that systematic errors due to e.g. the adopted distance modulus or the zero 
points of the metallicity scales are likely larger; however, these errors should manifest themselves 
as a constant offset in the adopted values for all stars and would only have a very minor 
impact on our discussion.

The ages we obtained from these fits are 10.1 Gyr and 10.7 Gyr for the b-SGB and f-SGB, respectively. 
For comparison, the age determined by Marin-Franch et al. (2009) is in the range between 9.8 Gyr and 
10.2 Gyr, depending on the method used. While our determination of age is quite inaccurate, 
because it depends on the value we adopted for the distance modulus, it is closely compatible 
with this value. This notwithstanding, the relative age difference of $\approx0.6$Gyr
between f-SGB and b-SGB stars remains constant provided that the unique difference in the 
chemical properties of the stars belonging to the two distinct SGBs is that related to the 
iron content (if there were any difference in the He and CNO abundances, this would 
affect the relative age dating as demonstrated by Cassisi et al. 2008 and Ventura et al. 2009).
This age difference is smaller than commonly assumed (Milone et al. 2008); this is 
because we assumed different [Fe/H] values for b- and f-SGB stars, as given by our analysis.

Microturbulence velocities {\it v$_t$} were obtained from surface gravities, using the calibration 
of Gratton et al. (1996), namely $v_t=0.322 \log{g} + 2.22$~km s$^{-1}$. They 
change by only small amounts within the gravity range of the programme stars (from 0.87 to 
1.03 km s$^{-1}$).

\begin{center}
\begin{figure}
\includegraphics[width=8.8cm]{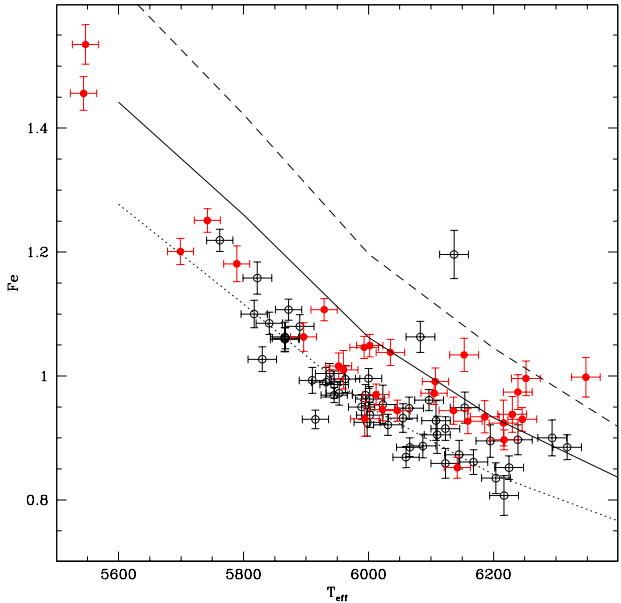}
\caption{Run of the parameter Fe, which represents the average intensity of 17 Fe I lines 
in our spectra, against effective temperature. Symbols are as in Fig.~\ref{f:fig1}. 
We overimpose three curves, representing the results obtained from a similar analysis of 
synthetic spectra computed with Fe abundances of [Fe/H]=-0.9, -1.1, and -1.3
(dashed, solid, and dotted lines, respectively). We note that the 
gravity and microturbulence of the models used for these synthetic spectra do not coincide 
with those appropriate for the individual stars, hence abundances cannot be derived by a 
simple interpolation in this plot.  }
\label{f:fig4}
\end{figure}
\end{center}

\begin{center}
\begin{figure}
\includegraphics[width=8.8cm]{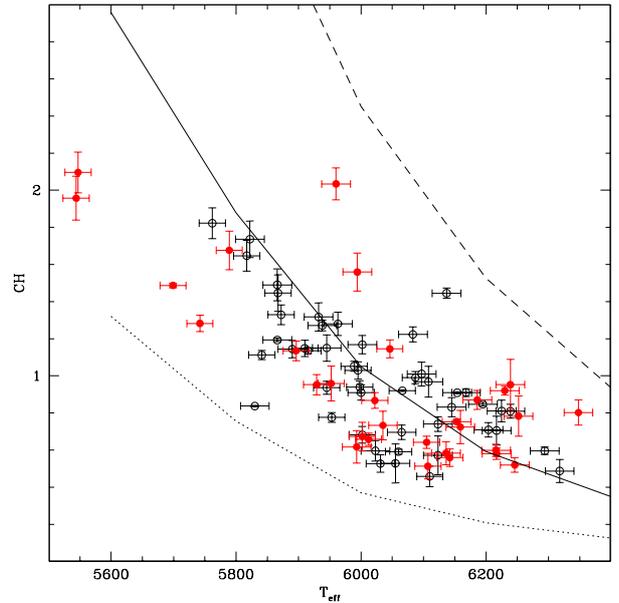}
\caption{Similar to Fig.~\ref{f:fig4}, but this time for the parameter CH, which represents 
the average strength of CH features within the G-band. The synthetic spectra were 
computed for [Fe/H]=-1.1, and [C/Fe]=-0.5, 0.0, and +0.5  (bottom to top line).}
\label{f:fig5}
\end{figure}
\end{center}

\begin{center}
\begin{figure*}
\includegraphics[width=18cm]{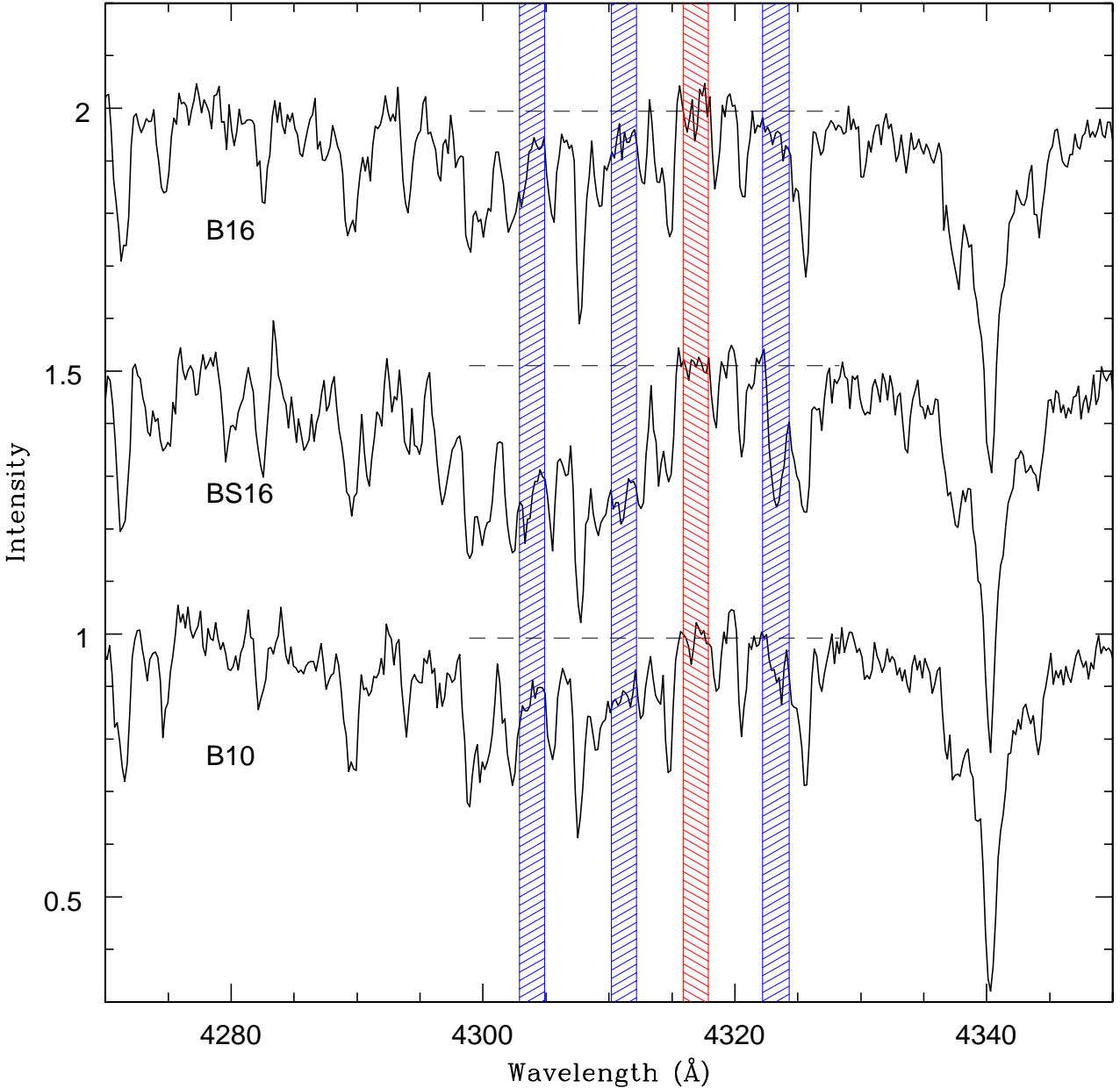}
\caption{Spectrum in the G-band region for the C-rich star BS16 ($T_{\rm eff}=5961$~K), 
compared with the spectra for a C-normal star B10 ($T_{\rm eff}=5940$~K) and C-poor star 
B16 ($T_{\rm eff}=5985$~K). Spectra have been offset vertically for a clearer display.
The spectral ranges used for the derivation of our CH index are shown as shaded areas, 
for easier comparison. The horizontal dashed line is at the average value within the
"continuum" reference band (4315.9-4317.9~\AA). }
\label{f:spectrach}
\end{figure*}
\end{center}

\subsection{Abundance analysis}

The second step consisted in measuring abundances from these spectra. There is virtually no 
unblended line in the spectra of the programme stars at the adopted resolution, and nowhere 
do the spectra ever reach a real continuum. The usual line analysis method 
was then not applicable. Our approach consisted in comparing the measures 
of fluxes in a number of spectral bands with similar measurements made on synthetic spectra. 
In general, we measured fluxes in both a narrow spectral region, dominated by a strong line, 
and a much wider region, used to normalize the spectra in a uniform way. We 
considered the following atomic lines: Al{\sc i} 3961.5 \AA; Mg{\sc i} 4057.5, 4167.3 \AA; Ca{\sc i} 
4226.7 \AA; Cr{\sc i} 4254.3, 4289.7 \AA; Mn{\sc i} 4030.7, 4033.1, 4034.5 \AA; Fe{\sc i} 4045.8, 4063.6, 
4071.7, 4132.1, 4143.9, 4187.0, 4202.0, 4235.9, 4250.1, 4260.5, 4271.2, 4307.9, 4325.8, 
4383.6, 4404.8, 4415.1, 4528.6 \AA; Sr{\sc ii} 4077.7, 4215.5 \AA; and Ba{\sc ii} 4554.0 Å. This list 
includes all lines in the useful spectral range stronger than 200~m\AA\ in the solar spectrum 
(save for the hydrogen lines, mentioned above, and the Ca II~H line, which is too strong to be
measured with the present approach). For all of these lines, 
we measured the flux within a band 1~\AA\ wide centred on the line, as well in a second band 
21~\AA\ wide, also centred on the line. The latter was used to provide a local reference for
the line index, which ensures that observed and synthetic spectra are normalized in the same 
way. Figure~\ref{f:fig3b} presents an example of the application of this procedure. We note that 
in this paper we only give results for Fe, Ca, Cr, Sr, and Ba lines.
The remaining lines give inconsistent results, and should be carefully examined for
possible errors in the adopted line lists. 

In addition to the atomic lines, we considered three spectral bands dominated by CH lines: 
4302.9-4304.9, 4310.2-4312.2, and 4322-4324.3~\AA. For these features, the local reference 
for the line index was obtained by considering the band 4315.9-4317.9~\AA, which is a relative high point 
in all our spectra. For examples of our spectra and the definition of these bands, we refer to
Figure~\ref{f:spectrach}.

The synthetic spectra we computed are based on the Kurucz (1993) set of model atmospheres (with 
the overshooting option switched off), and line lists from Kurucz (1993) CD-ROM's, using our own 
synthesis code; we checked that the parameters for the strong lines contained in these lists 
were equal to the updated values from the VALD database (Kupka et al. 2000)\footnote{See URL vald.astro.univie.ac.at}. 
The adopted solar abundances were 8.67, 6.34, 5.71, 7.63, 3.06, and 2.28 for
C, Ca, Cr, Fe, Sr, and Ba, respectively, in the usual spectroscopic scale where $\log{\rm n(H)}=12.0$.
There are offsets with respect to other sets of solar abundances; for instance they are
up to 0.2 dex higher than those listed by Asplund et al. (2009). While some of these
offsets should be applied when using different sets of model atmospheres, we underline that
we are mainly interested in star-to-star abundance differences within NGC~1851, so that global offsets 
in the solar abundances do not affect any of the main conclusions of this paper. 
Synthetic spectra were computed for [A/H]=-1.6, -1.3, -1.1, -0.9 and -0.6, save for Ba, for which we 
adopted [A/H]=-1.0, -0.7, -0.5, -0.3, and 0.0; in this way, we virtually never need to extrapolate 
beyond the computed grid. The spectra were then convolved 
with a Gaussian having a FWHM of 0.5~\AA\ that closely mimics the instrumental profile.

Our derivation of abundances from measured fluxes involved several steps. We 
wished to compute only a small set of synthetic spectra. In addition, we wished to 
estimates the internal errors by placing the measures for different lines on a uniform scale. In 
practice, this was done by dividing the integral of the residual profile measured for each line 
$(1-F_\lambda)$\ by its average value among all programme stars. This quantity is of the order 
of unity, and has a similar run with temperature for all lines of a given element
because the excitation potentials of the lines are quite similar to each other, often because they 
belong to the same multiplet. We then averaged results for different lines of a given element; 
the r.m.s scatter in the individual values provides an estimate of the internal errors, which 
can then be used to compare abundances obtained for different stars. In practice, slightly better 
results were obtained by weighting the individual lines according to the scatter in the residuals 
with respect to the mean values. These average quantities were then compared with the results 
obtained by applying the same procedure to the synthetic spectra. Fig.~\ref{f:fig4} shows the results of 
the application of this procedure for Fe lines (we define as Fe the parameter describing the 
average intensity of the 17 Fe I lines that we measured on our spectra).

Fig.~\ref{f:fig5} shows similar results, this time for the parameter CH obtained considering the 
spectral regions close to the head of the G-band.

Abundances of the various elements in the individual stars may then be obtained by interpolations 
within the graphs. Rather than using interpolations, we however preferred to evaluate low order 
polynomials in $T_{\rm eff}$\ and the various indices by least squares fitting, and then correct 
for gravity and microturbulence velocity appropriate for individual stars, using the sensitivities 
of Table~\ref{t:tab2}. The abundances we obtained by this process are listed in Tables~\ref{t:tab1a} 
and ~\ref{t:tab1b}. For those cases in which several features were used, we also provide an internal 
error, which is a statistical error, and only includes the contribution due to the scatter between 
results obtained for different lines. Table~\ref{t:tab2} gives the sensitivity of the abundances we 
obtain for various elements to the the atmospheric parameters. We also list the total 
uncertainty, which was obtained by assuming errors of 30~K, 0.03 dex, and 0.2 km s$^{-1}$\ in effective 
temperature, surface gravity, and microturbulence velocity, respectively. Systematic errors are much 
larger, of the order of 0.1~dex or more. However, insofar as they are similar among all programme stars, 
they do not affect our discussion.

\begin{table}[htb]
\centering
\caption[]{Sensitivity of abundances to the adopted atmospheric parameters.}
\begin{tabular}{lccccc}
\hline
Parameter &	$T_{\rm eff}$ &	$\log{g}$ & $[$A/H$]$ & $v_t$ & Total \\
\hline
Variation & +200 K & +0.3 dex & +0.1 dex & +0.3 km s$^{-1}$\ &    \\	
Error	  & 30 K   & 0.03 dex &          & 0.2 km s$^{-1}$\  &    \\	
\hline
$[$Fe/H$]$  & +0.22	& -0.18	&  & -0.04	& 0.05 \\
$[$Ca/Fe$]$	& ~0.00	& +0.02	&  & +0.02	& 0.01 \\
$[$Cr/Fe$]$	& +0.14	& -0.05	&  & -0.08	& 0.06 \\
$[$Sr/Fe$]$	& -0.13	& +0.18	&  & -0.03	& 0.03 \\
$[$Ba/Fe$]$	& +0.05	& +0.04	&  & -0.09	& 0.05 \\
$[$C/Fe$]$	& +0.03	& +0.09	&  & +0.03	& 0.02 \\
\hline
$\Delta(u-y)$ & +0.078 & +0.025 & -0.023 & -0.016 & 0.021 \\
$[$N/Fe$]$	  & +0.98 &  +0.31 & -0.29  &  -0.19  & 0.26 \\
\hline
\end{tabular}
\label{t:tab2}
\end{table}

\section{Abundances for individual elements}

\begin{center}
\begin{figure}
\includegraphics[width=8.8cm]{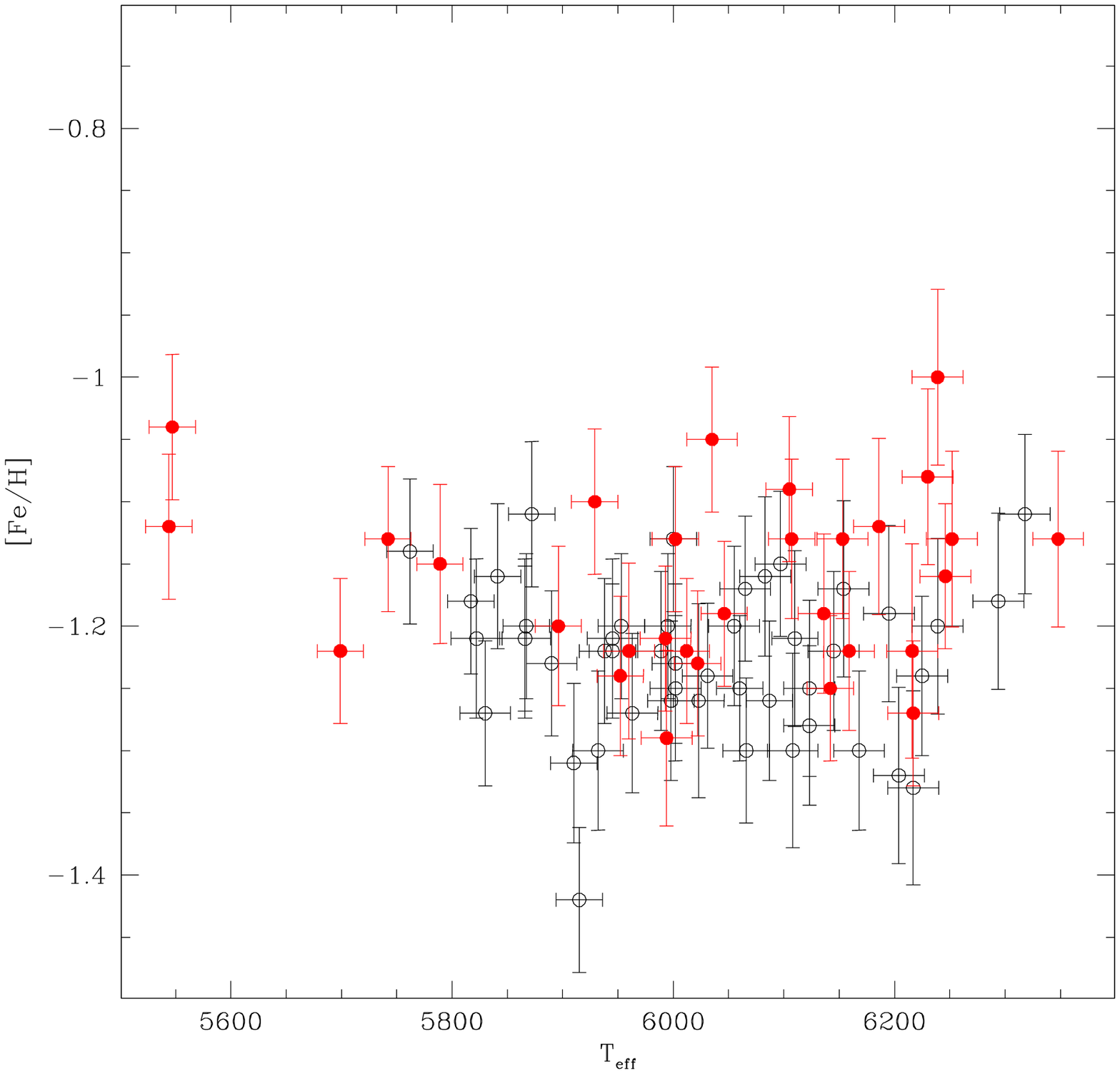}
\caption{Run of the abundance of Fe with effective temperature. Symbols are similar to those in Fig.~\ref{f:fig1} }
\label{f:fig6}
\end{figure}
\end{center}

\begin{center}
\begin{figure}
\includegraphics[width=8.8cm]{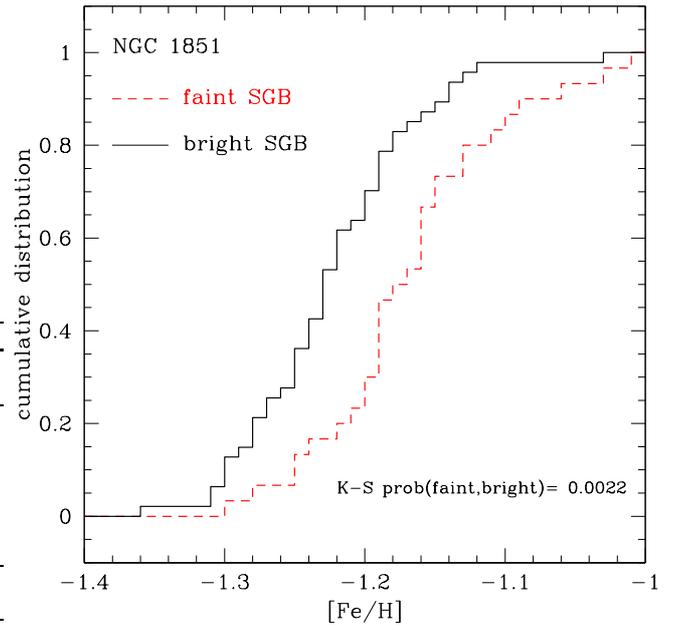}
\caption{Comparison between the cumulative distribution of [Fe/H] values for b-SGB and 
f-SGB stars. We also show the result of the application of the Kolmogorov-Smirnov test 
to determine the probability that the two distributions were extracted
from the same parent population. }
\label{f:figdist}
\end{figure}
\end{center}

\subsection{Iron}

Fig.~\ref{f:fig6} gives the run of the abundance of Fe with the effective temperature. 
No trend with temperature is present. On average, 
we obtain [Fe/H]=$-1.20\pm 0.01$\ (r.m.s.=0.07~dex). This value compares very well with
recent determinations of the metal abundance of NGC~1851, as summarized in 
Table~\ref{t:tab3}.

\begin{table*}[htb]
\centering
\caption[]{Fe abundance in NGC~1851 for different groups of stars}
\begin{tabular}{lcccc}
\hline
Group     &	N. stars&[Fe/H]	&r.m.s.& paper \\
\hline
SGB       &  77 & $-1.20\pm 0.01$ & 0.07 & This paper \\
SGB-b     &  47 & $-1.23\pm 0.01$ & 0.06 & This paper \\
SGB-f     &  30 & $-1.16\pm 0.01$ & 0.07 & This paper \\
Upper RGB &  13 & $-1.18\pm 0.02$ & 0.07 & Carretta et al. (2010c) \\
Upper RGB & 121 & $-1.15\pm 0.01$ & 0.05 & Carretta et al. (2011a) \\
RHB       &  55 & $-1.14\pm 0.01$ & 0.06 & Gratton et al. (2012b) \\
Lower RGB &  13 & $-1.18\pm 0.03$ & 0.11 & Gratton et al. (2012b) \\
\hline
\end{tabular}
\label{t:tab3}
\end{table*}
 
All of these sets of abundances were obtained using the Alonso et al. (1996, 
1999) temperature scales (the BASTI one in the present paper), the Kurucz (1992) model 
atmospheres, and local thermodynamic equilibrium. The [Fe/H] value we get for the b-SGB is lower (by 
0.08~dex) than the value obtained for the RHB, in spite of their being 
likely to belong to the same population. This difference is much larger than the 
statistical errors (those listed in Table~\ref{t:tab3}) and might perhaps agree 
with expectations for the impact of sedimentation. However, we are not inclined to
attribute much importance to this difference, given the potential offsets that 
can be present for these different analyses. We propose instead that this comparison 
may give the reader a clearer idea of the uncertainties that exist in the zero points of 
these different abundance determinations other than the statistical errors.

Stars on the f-SGB have systematically higher Fe abundances than those on the 
b-SGB. On average, we have [Fe/H]=$-1.227\pm 0.009$\ for the b-SGB stars, and 
[Fe/H]=$-1.162\pm 0.013$\ for the f-SGB. The r.m.s. scatter in the abundances of 
0.062 dex and 0.071 dex, respectively, can be attributed to a combination of the effects of 
the line-to-line scatter ($\sim 0.04$~dex: see Tables~\ref{t:tab1a} and ~\ref{t:tab1b}) 
and the internal errors in the atmospheric parameters ($\sim 0.05$~dex, see 
Table~\ref{t:tab2}). The difference between the average Fe abundances of the b-SGB and 
f-SGB is significant at more than 4$\sigma$: it cannot be attributed to a random 
fluctuation. This is clearly shown by Figure~\ref{f:figdist}, where we compared
the cumulative distributions of [Fe/H] values for b-SGB and f-SGB stars. An
application of the Kolmogorov-Smirnov test to determine the probability that the two 
distributions were extracted from the same parent population returns a very low 
value of 0.0022. In addition, the difference cannot be explained by errors in the adopted masses:
to eliminate this difference, we would have to assume that the mass of f-SGB stars is ~20\% 
higher than the value for the b-SGB stars, which does not sound reasonable (if 
anything, we would expect that the masses of f-SGB stars are lower, not higher than those 
of b-SGB stars, and in any case the difference should be very small, $<2$\%). We are then 
inclined to conclude that there is a real difference between the metal abundances 
of f-SGB and b-SGB. We note that Carretta et al. (2010c, 2011a) also concluded that there is a 
real range of metal abundances in NGC~1851, while Villanova et al. (2010) did not 
find any difference when stars were divided into groups according to their Str\"omgren 
indices. This might be understood if - for the upper RGB - the subdivision of
stars into groups given by Str\"omgren photometry did not exactly reproduce that
in b-SGB/f-SGB.

We note that there appear to be a couple of outliers in Figure~\ref{f:fig4}.
The most discrepant case is star AS57, which we actually omitted from Table~\ref{t:tab1a}. 
Its Fe index would correspond to [Fe/H]=-0.72. Inspection of the spectrum confirms that 
this star has stronger lines than the other stars of similar $T_{\rm eff}$. In the 
theoretical $T_{\rm eff}$-gravity diagram of Figure~\ref{f:fig3}, this star 
($T_{\rm eff}$=6137, $\log{g}$=3.936) corresponds to the most discrepant data point having too low a gravity 
for its adopted $T_{\rm eff}$. Our assumed $T_{\rm eff}$ might indeed be overestimated; if we adopt 
$T_{\rm eff}$ from $(B-I)$\ colour ($T_{\rm eff}$=6004 K), we have $\log{g}$=3.895, in 
agreement with the values inferred for other stars of similar $T_{\rm eff}$. However, the [Fe/H] value ([Fe/H]=-0.89) 
would still differ by more than five standard deviations from the values for other b-SGB 
stars and again a direct comparison shows that the observed iron lines are much stronger than for other stars 
of even this lower temperature. The $T_{\rm eff}$ that would be required to recover a typical [Fe/H] value is 
$\sim 5700$~K, but this would be excluded by both the spectrum and the photometry. An alternative 
possibility is that this star is actually a binary (either real or apparent).

The next possible outlier in Figure~\ref{f:fig4} is star BS19, which is the warmest in 
our sample. However, in this case there is no real anomaly, and the apparent discrepancy is 
simply caused by the adopted values for $\log{g}$ and $v_t$ for this
star which considerably differ from those used to draw the lines in Figure~\ref{f:fig4}.

\begin{center}
\begin{figure}
\includegraphics[width=8.8cm]{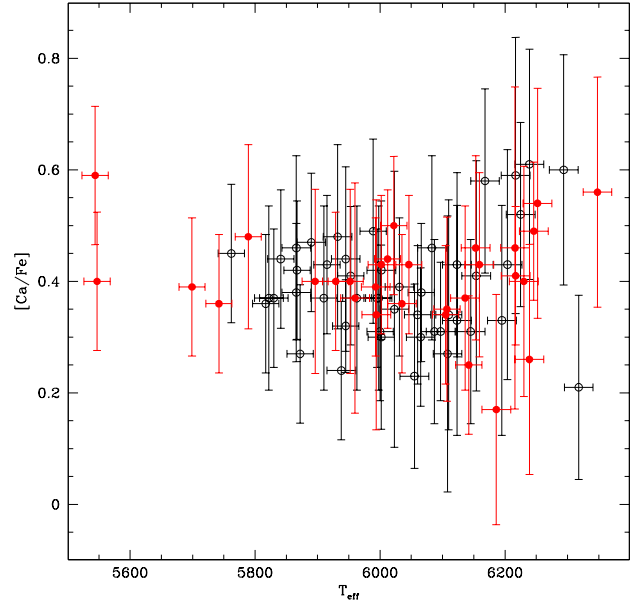}
\caption{Run of the [Ca/Fe] abundance ratio with effective temperature. Symbols 
are similar to those in Fig.~\ref{f:fig1} }
\label{f:fig7}
\end{figure}
\end{center}

\begin{center}
\begin{figure}
\includegraphics[width=8.8cm]{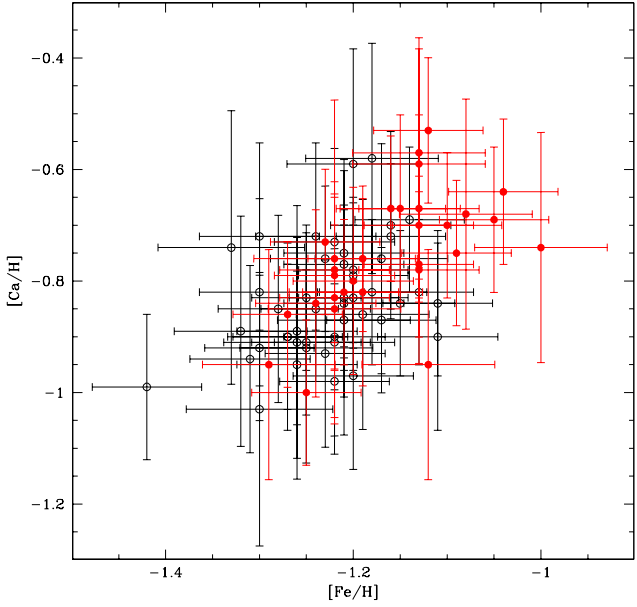}
\caption{Comparison between [Ca/H] and [Fe/H]. Symbols are similar to those in Fig.~\ref{f:fig1} }
\label{f:fig8}
\end{figure}
\end{center}

\subsection{Calcium}

Fig.~\ref{f:fig7} shows the run of the [Ca/Fe] ratio with effective temperature. This 
run is uniform and flat at [Ca/Fe]=$0.40\pm 0.01$, with an r.m.s.=0.09 dex. While this result 
is expected to be almost independent of the choice of the atmospheric parameters, 
the scatter still appears remarkably small for an abundance based on a single line in 
moderately low resolution spectra. The average value is similar to those derived by 
Carretta et al. (2010c, 2011a: [Ca/Fe]=$0.30\pm 0.02$) and Gratton et al. (2012b: 
[Ca/Fe]=0.42) from an analysis of RGB and RHB stars, respectively.

Fig.~\ref{f:fig8} compares the abundances of Fe and Ca. This figure clearly shows that 
both Fe and Ca are more abundant in f-SGB than in b-SGB stars. We found average abundances 
of [Ca/H]=$-0.836\pm 0.014$\ for b-SGB stars, and [Ca/H]=$-0.756\pm 0.020$\ for f-SGB 
stars. This implies that large differences in Ca/Fe would be unable to explain the spread 
in $Cauvby$\ photometry noted by Lee et al. (2009). This agrees with the discussions of 
Carretta et al. (2010b) and Sbordone et al. (2011).

\begin{center}
\begin{figure}
\includegraphics[width=8.8cm]{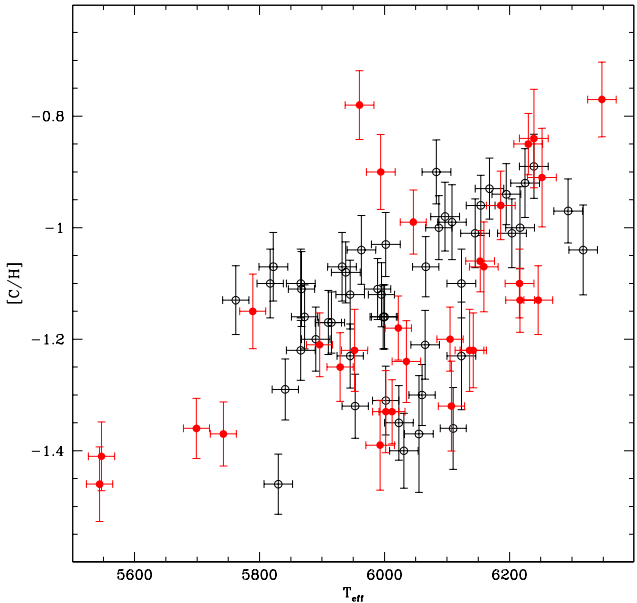}
\caption{[C/H] ratio as a function of effective temperature. Symbols are similar to those in Fig.~\ref{f:fig1} }
\label{f:fig9}
\end{figure}
\end{center}

\subsection{Carbon}

Fig~\ref{f:fig9} displays the run of [C/H] with effective temperature. This plot is 
very instructive and is worthy a closer examination. We begin by considering the b-SGB star. 
First, we note that there is a clear overall trend for decreasing C abundances with decreasing 
temperature, with a slope of about 0.0005 dex/K. This small trend could possibly be
due to the deepening of the convective envelope when the star moves from the main-sequence 
turn-off to the base of the RGB. Furthermore, the distribution appears 
skewed, with a group of 27 stars that define a very narrow relation (r.m.s. of 
0.033~dex, excluding one star that has a slightly higher than average C abundance) at the 
upper envelope of the distribution, the remaining 19 stars scattering below, within a 
range of about 0.4~dex. We might expect that the C-normal group are N-poor, and the 
C-poor stars are N-rich; unluckily, the CN band at 4216~\AA\ is too weak to confirm 
this difference. The distribution of f-SGB stars appears to be more scattered, the stars
being on average a little poorer in C than the b-SGB stars. However, this large scatter may
be caused by the stars of C-normal and C-poor groups having different distributions,
rather than a general property of all stars. This can be shown by comparing the offsets in 
C abundances from a mean line representing the bulk of b-SGB stars, as represented by the equation:
\begin{equation}
{\rm offset[C/H]} = {\rm [C/H]} -– [0.0005~(T_{\rm eff}  - 6200) - 0.965]
\end{equation}
We may then separate C-normal stars from C-poor stars at a [C/H] value that is that given
by the relation minus twice the r.m.s. of the bulk of 
b-SGB stars around the relation itself; that is, C-normal stars have offset[C/H]$>-0.067$, 
while C-poor stars have offset[C/H]$<-0.067$. Among b-SGB stars, there are 27 C-normal stars, 
17 C-poor stars, and 1 star slightly more C-rich than the C-normal ones. Among the f-SGB 
stars, there are 22 C-poor stars, 3 C-normal stars, and 5 C-rich stars. The very different 
incidences of C-rich, C-normal, and C-poor stars among the two populations is remarkable.

As mentioned above, the spectra of a few stars contain a G-band that is definitely stronger 
than in the remaining spectra;
an example is shown in Figure~\ref{f:spectrach}, where we compare a spectrum of the f-SGB
star BS16 with that of two other f-SGB stars of similar temperature (B10 and B16). There
is little doubt that the G-band is indeed stronger in the first star. The C abundances we
derived using our procedure are [C/Fe]=$+0.43\pm 0.03$\ for BS16, $0.02\pm 0.05$\ for
B10, and $-0.20\pm 0.05$\ for B16. In our opinion, these C-rich stars are most likely 
the progenitors of the stars on the anomalous RGB branch in the $v,(v-y)$ diagram, 
for the reasons that we discuss below.

Finally, Lardo et al. (2012) analysed the C and N contents of a 
quite large number of stars close to the subgiant branch of NGC~1851. A comparison
with their results is needed. However, not only are there only three stars in
common between the samples, but we also note that these authors attributed an effective 
temperature of about 5850~K to the turn-off stars of NGC~1851. This is about 500 degrees 
cooler than expected for the BASTI group isochrones (Pietrinferni et al. 2006) they used 
elsewhere in their paper. The main 
effect of such a low temperature scale on their analysis is to severely underestimate 
the abundances of C, which are indeed very low with typical values of [C/Fe]$\sim -0.6$. 
In turn, this yields too low C+N abundances for the C-normal, N-poor stars, which are 
predominant on the b-SGB but are only a minority of the f-SGB stars (see Section 4.3), 
and leads to the authors' conclusion about the difference between the sum of C+N abundances 
for the two branches. Furthermore, the large scatter that is clearly evident in a star-to-star 
comparison among the only three stars in common between the two samples induces us to 
suspect that they underestimated the impact of their observational errors on their C abundances. 
Since N abundances are derived from CN bands, any error in the C abundances propagates 
into the N ones, creating a spurious C-N anticorrelation.

\begin{center}
\begin{figure}
\includegraphics[width=8.8cm]{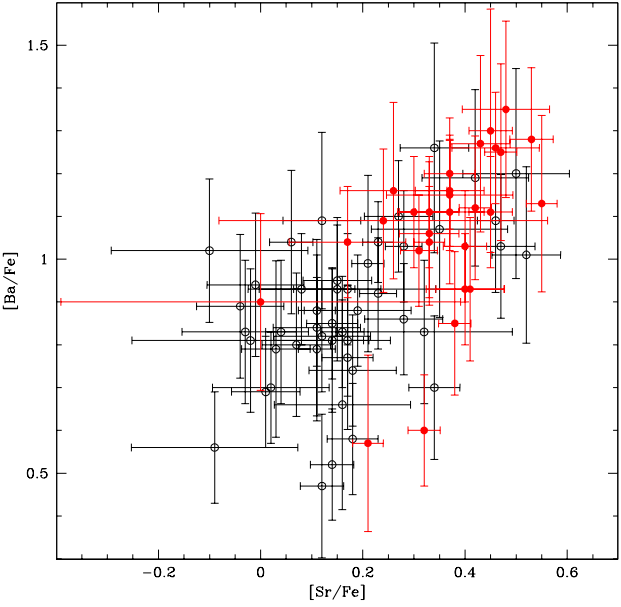}
\caption{Comparison between [Sr/Fe] and [Ba/Fe] abundance ratios. Symbols are similar to those in Fig.~\ref{f:fig1} }
\label{f:fig10}
\end{figure}
\end{center}

%\begin{center}
%\begin{figure}
%\includegraphics[width=8.8cm]{NGC1851sg_srbah.ps}
%\caption{Comparison between [Sr/H] and [Ba/H] abundance ratios.  Symbols are as in Fig.~\ref{f:fig1} }
%\label{f:fig11}
%\end{figure}
%\end{center}

\begin{center}
\begin{figure}
\includegraphics[width=8.8cm]{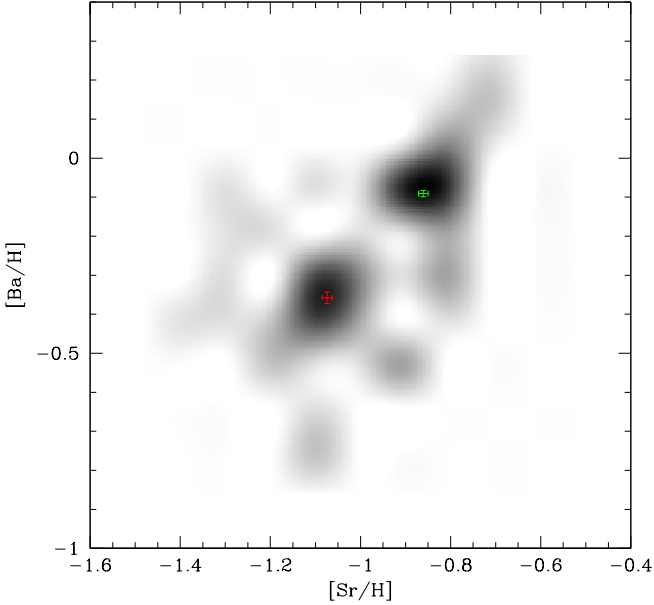}
\caption{Hess diagram constructed for the  [Sr/H] and [Ba/H] abundance ratios.
Crosses indicate the mean Sr and Ba contents of the two bumps with their internal errors. }
\label{f:fig11b}
\end{figure}
\end{center}

\begin{center}
\begin{figure}
\includegraphics[width=8.8cm]{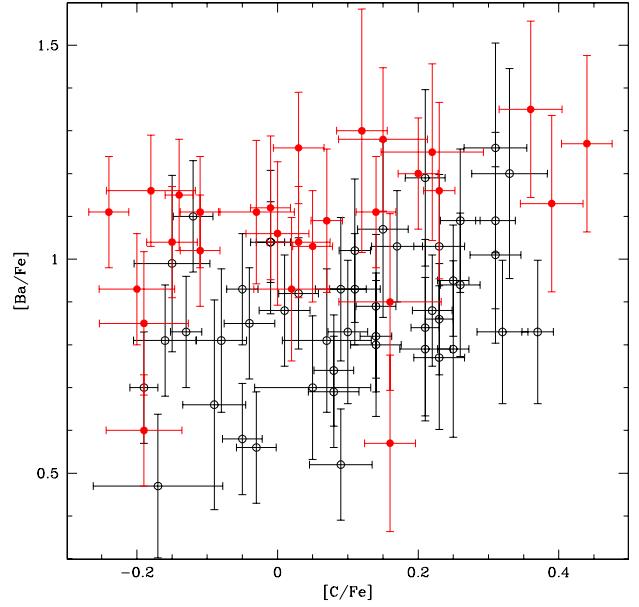}
\caption{Correlation between the Ba and C abundances. Symbols are similar to those in Fig.~\ref{f:fig1} }
\label{f:fig12}
\end{figure}
\end{center}

\begin{center}
\begin{figure}
\includegraphics[width=8.8cm]{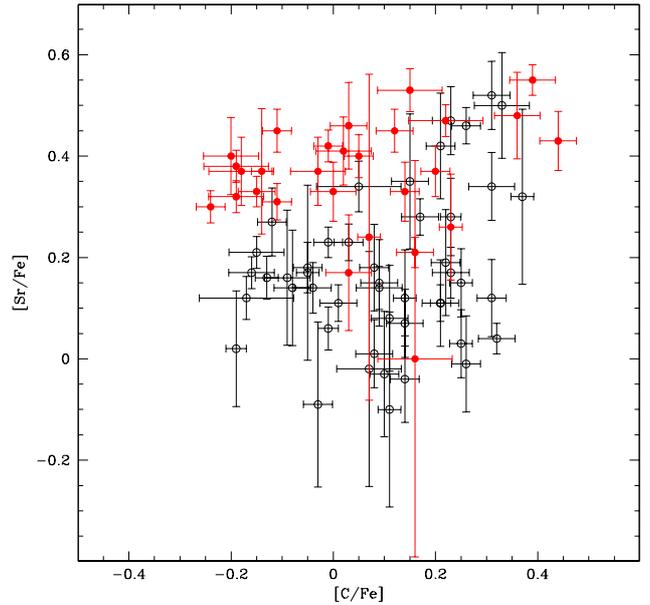}
\caption{Correlation between Sr and C abundances. Symbols are similar to those in Fig.~\ref{f:fig1} }
\label{f:fig13}
\end{figure}
\end{center}

\subsection{Neutron-capture elements}

We measured abundances for two n-capture elements (Sr and Ba). Both elements are mainly 
produced by the s-process in the solar system. Fig.~\ref{f:fig10} and \ref{f:fig11b} 
compare the abundances of Sr and Ba. These plots show two interesting features:
\begin{itemize}
\item Both Sr and Ba are more abundant in f-SGB than b-SGB stars. The differences are 
$\Delta$[Sr/Fe]=$0.19\pm 0.03$~dex and $\Delta$[Ba/Fe]=$0.20\pm 0.04$~dex. Differences are 
even larger in [Sr/H] and [Ba/H] ($0.25\pm 0.03$\ and $0.26\pm 0.04$, respectively).
\item In both groups, there is a clear correlation between [Sr/Fe] and [Ba/Fe], suggesting 
that there is a real spread in the abundances of n-capture elements. This agrees with the
correlation  found for RGB and RHB stars, if we identify RHB stars with b-SGB ones.
\item Both Sr and Ba have bimodal distributions as previously noticed by Villanova et al. (2010).
This bimodality is visible in both Figures~\ref{f:fig10} but is more clearly evident in 
Fig.~\ref{f:fig11b}, where we present a Hess diagram of the [Ba/H] versus [Sr/H] relation. Two 
clear bumps are visible. Two crosses indicate the mean Sr and Ba content of the two bumps 
with their internal errors. The bimodality of $s-$elements (both light and heavy) in NGC~1851 
obtained by Villanova et al. (2010) appears to be definitely proven, although there are
offsets between the mean abundances of the two groups of stars that may depend on the differences
between the analysis methods.
\item If we consider the distribution of stars in the SGB (Fig.~\ref{f:fig0}), the
[C/H] vs T$_{\rm eff}$ diagram (Fig.~\ref{f:fig9}), and the
[Ba/H] vs. [Sr/H] diagram (Fig.~\ref{f:fig11b}), we see that there is no one-to-one 
relation among these quantities. For example, both branches are populated by C-normal and 
C-poor stars, while not all stars that are Ba-rich are also C-poor and viceversa. 
Measurement errors are clearly responsible for part of these perplexing results, but 
the lack of correlation can be hardly entirely justified in this way.
\end{itemize}

We then looked for a correlation between n-capture elements and C abundances (see 
Fig.~\ref{f:fig12} and \ref{f:fig13}). There is some correlation for Ba, while the result 
for Sr is much more dubious. However, both results might be influenced by the small trends 
with effective temperatures.

Finally, we note that Sr and Ba abundances for the few C-rich stars are slightly higher 
than for the other stars, but the difference is not large. Typical values are 
[Sr/Fe]$\sim 0.45$\ and [Ba/Fe]$\sim 1.25$, that are $\sim 0.3$~dex higher than average.

\section{Nitrogen abundances}

Clearly, CNO abundances play a very special role in NGC~1851. We now examine the 
evidences for N abundances determined for the stars we observed. Unfortunately, 
we do not have any N abundance indicator in our spectra because the CN band at 
4216~\AA\ is too weak in the programme stars. In these warm stars, the best 
indicator of N is the NH band. While we have no spectroscopic data at the 
relevant wavelengths, we have Str\"omgren photometry for most of the stars.
We tried several combinations of the Str\"omgren indices and found that best 
results are obtained using $u-y$, where the $u$-band is affected by NH and CN 
lines, which are not present in the $y$-band. Since $u-y$\ is both temperature and 
gravity dependent, we should use the offset from an average relation of this 
index against temperature. We called this offset $\Delta(u-y)$, whose definition 
is:
\begin{equation}
\Delta(u-y) = (u-y) - (0.0005708~T_{\rm eff}  - 5.056).
\end{equation}
Values for individual stars are given in the last column of Table~\ref{t:tab1a}.

As an exercise, we then synthesized Str\"omgren colours for subgiants with 
different N-contents. We adopted a similar approach to Carretta et al. (2011b;
see also Sbordone et al. 2011 for a similar discussion). We found that
a huge excess of N, say [N/Fe]=1.5 (2.0), shifts the $u-y$\ colour by as much 
as 0.10 (0.16) mag, with virtually no effects on all bands apart from $u$ (the test 
was done for $T_{\rm eff}=6000$~K, which is typical of the programme 
stars). Of course, $\Delta(u-y)$\ also depends on other parameters 
(temperature, gravity, Fe abundance, and microturbulence velocity). We explored 
this dependence by modifying these parameters in the synthesis, and estimating 
the impact on $\Delta(u-y)$. The relevant values are given in Table~\ref{t:tab2},
in terms of the effects of variation of the parameters on both $\Delta(u-y)$\ and 
the [N/Fe] values that we would 
infer from these data. While the sensitivities are not at all negligible, the 
accuracy of our determination of the atmospheric parameters ensures that the 
internal errors in $\Delta(u-y)$\ and [N/Fe] are dominated by the photometric 
errors.

\begin{center}
\begin{figure}
\includegraphics[width=8.8cm]{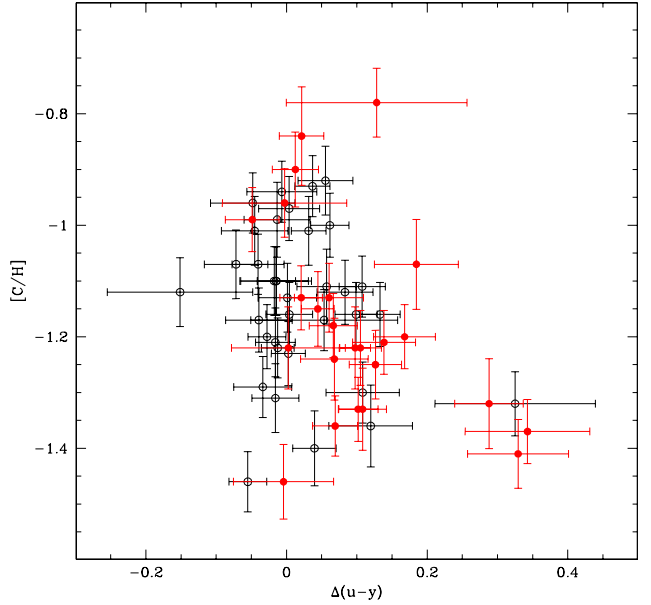}
\caption{Correlation between $\Delta(u-y)$\ and C abundances. Symbols are similar to those in Fig.~\ref{f:fig1} }
\label{f:fig14}
\end{figure}
\end{center}

\begin{center}
\begin{figure}
\includegraphics[width=8.8cm]{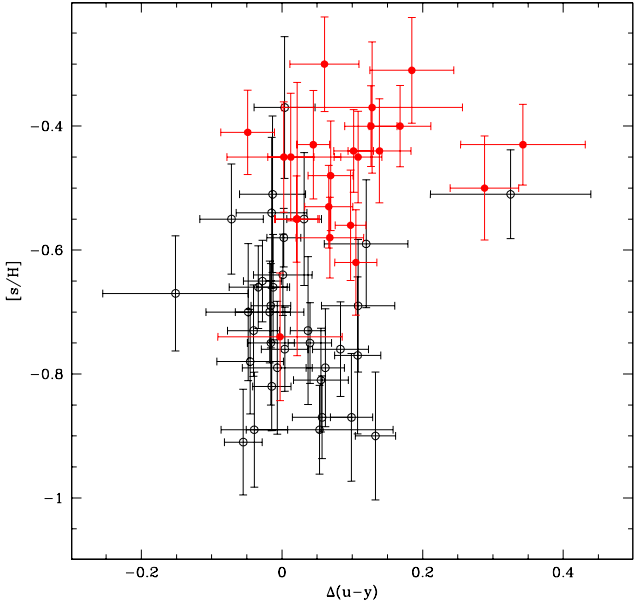}
\caption{Correlation between $\Delta(u-y)$\ and the average of Sr and Ba abundances. Symbols are similar to those in Fig.~\ref{f:fig1} }
\label{f:fig15}
\end{figure}
\end{center}

In Figures~\ref{f:fig14} and \ref{f:fig15}, we then compared the $\Delta(u-y)$\ with 
[C/H] and [s/H]=[(Sr+Ba)/H] abundances. Although the scatter is quite large, owing to the 
photometric errors, $\Delta(u-y)$\ is positively correlated with [s/H] and anticorrelated
with [C/H] for the f-SGB stars, while the results are less clear for b-SGB ones. This is
reasonable because $\Delta(u-y)$\ may be considered as a measure of the N abundance 
(even though with large errors). To quantify these effects, we performed various tests.

We first estimated the average values of $\Delta(u-y)$\ in both groups of SGB stars. We
obtained values of $0.021\pm 0.013$~mag (35 stars, r.m.s.=0.078 mag) and $0.101\pm 0.021$~mag 
(24 stars, r.m.s.=0.103 mag) for the b-SGB and f-SGB stars, respectively. There is then a
systematic difference of $0.080\pm 0.024$~mag between the two groups, which is significant
at more than the 3-$\sigma$\ level. This offset is much larger than expected for the
tiny differences in gravity and metal abundances (which can justify only about one fifth of the 
difference), and should then be real and likely related to a difference in the average N 
content. We would then expect f-SGB stars to have on average a [N/Fe] which is $\sim 1.0$~dex higher
than that of b-SGB stars. Since we find a large difference in the fraction of C-poor stars, 
this higher abundance is 
likely due to a much larger incidence of N-rich stars. This closely agrees with the lower 
average [C/Fe] values obtained for the f-SGB stars, in the framework of an anticorrelation 
between C and N abundances.

Second, such an anticorrelation is evident when directly examining the results for
individual stars. We estimated the linear correlation coefficient between [C/H] and
$\Delta(u-y)$, and found a value of r=0.331 (59 stars, $<0.005$ probability to be a
random result) when considering all stars. The correlation coefficient is insignificant 
for the b-SGB stars alone (r=0.136 over 35 stars), while it is quite high (r=0.458 over
24 stars) for f-SGB stars alone ($\sim 0.02$ probability of being a random result).

We conclude that there is a significant anticorrelation between [C/H] and
$\Delta(u-y)$, which is mainly driven by the f-SGB stars. These results may indicate
that most b-SGB stars are C-normal and N-normal; while the majority of 
f-SGB ones are C-poor and N-rich. 

We note that there is a small group of four (three on the f-SGB and one 
on the b-SGB) stars that have very large values of $\Delta(u-y)\sim 0.3$. Given their 
high N abundances, these stars might be considered potential
progenitors of the stars on the anomalous RGB branch in the $v,(v-y)$ diagram 
instead of the C-rich stars mentioned above. However, we first note that with 
an average [Ba/Fe]=0.32 these stars do not have a large overabundance of Ba,  which is
an important characteristic of the stars on the anomalous RGB branch (Villanova et al. 2010; 
Carretta et al. 2011b). Secondly, such a large $\Delta(u-y)$ value, if naively 
interpreted in terms of anomalous N overabundances, would produce a huge value of 
[N/Fe]$>>2$. We deem that it is more likely that the large values of $\Delta(u-y)$\ found for 
these stars is due to a combination of high (but not exceptional) N abundances with 
observational errors (errors shown in Figures~\ref{f:fig14} and \ref{f:fig15}
are the internal errors in the photometry, and may underestimate real errors in case
of blends). Ultraviolet spectra of these stars would of course be extremely helpful.

We may compare this result with that of Lardo et al. (2012) mentioned in the introduction.
Unsurprisingly, we found a very good correlation between $\Delta(u-y)$\
used throughout this paper and the residuals around the mean relation between the
index used by Lardo et al. ($(u-y)+(v-b)$) and V magnitudes for the SGB.

Finally, we note that C-rich stars have small values of $\Delta(u-y)$, which are
indicative of normal (low) N abundances. Their composition suggests that they formed from material 
polluted by the products of triple-$\alpha$\ reactions.

\section{Discussion and conclusions}

A short summary of the results of our analysis is as follows:
\begin{itemize}
\item We have been able to clearly distinguish between the b-SGB and f-SGB populations of NGC~1851 in the 
theoretical $T_{\rm eff} - \log{g}$\ diagram.
\item We have found a small but significant difference in the metal abundances. The b-SGB has 
[Fe/H]=$-1.23\pm 0.01$, while for the f-SGB we obtained [Fe/H]=$-1.16\pm 0.01$. Hence, the f-SGB 
is slightly more metal-rich than the b-SGB. The sense of this difference is the same as that 
found by Marino et al. (2012b) for M~22.
\item If we then assumed the same He and CNO/Fe ratios for the two populations, the difference in
age between them is reduced to 0.6 Gyr only. The analysis of the HB performed in Gratton et al. (2012b)
shows that the BHB (which is likely mostly populated by the progeny of the f-SGB population) is slightly 
more He-rich (Y$\sim $0.29 vs. Y$\sim $0.25) than the RHB (which is likely the progeny of the b-SGB
population). However, detailed comparisons with isochrones computed for this purpose shows that this 
conclusion about the age difference is unaffected by modifications of the He abundances
(which however affect the masses of the stars). On the other hand, CNO abundances are much more 
a concern. If for instance we assume that the sum of C+N+O is larger by a factor of 2 for f-SGB stars
than b-SGB stars, then the first would be found to be younger than the second by $\sim 0.4$~Gyr, which is 
almost exactly the opposite of what we get by assuming the same CNO/Fe ratio. This would of course
completely change the evolutionary scenario appropriate to interpreting our observations.
\item The variation in Fe abundances is paired by abundances in other elements such as Ca, that
is the [Ca/Fe] ratio is very similar in the two populations.
\item The neutron-capture elements Sr and Ba show much larger differences between the b-SGB and
the f-SGB than Fe or Ca. These elements have a strong bimodality, as found by Villanova et al.
(2010). This result is again similar to that found by Marino et al. (2012b) for M~22.
\item Both the b-SGB and the f-SGB stars exhibit a spread in C abundances (this result coincides
with that of Lardo et al. 2012), but the ratio of C-normal to C-poor stars are very different
in the two populations: they are 27:17 and 3:22 for the b-SGB and f-SGB, respectively. 
\item We were unable to derive N abundances from our spectra. When we assumed that an index $\Delta(u-y)$\
derived from Str\"omgren photometry provides information about the N abundance, we found that there is a quite
good C-N anticorrelation for the f-SGB, while results are unclear for the b-SGB. In addition, we found
that on average the f-SGB stars are more N-rich than b-SGB stars. This is likely related 
to the very different incidence of C-normal and C-poor stars in the two groups.
\item There are a few abnormally C-rich stars, most of them (five out of six) in the f-SGB. The C 
overabundance of these stars is obvious even from a simple inspection of the spectra. These stars 
are also Sr- and Ba-rich, though there appears to be other stars in the cluster with comparable
Sr and Ba abundances. We propose that these stars are the most likely progenitors of the stars 
on the anomalous RGB in the $v-(v-y)$\ diagram, since (i) a high C abundance might be an
explanation for the anomalous $v-y$\ colours (Carretta et al. 2010c), (ii) these stars are
Ba-rich, and (iii) the number of observed C-rich stars (6 out of 78) is in reasonable agreement
with the fraction of RGB stars on the anomalous RGB. We note that the N-abundance 
of the C-rich stars is similar to the average for the other stars, suggesting that they owe their 
large C-abundance to triple-$\alpha$\ reactions in the progenitor population and not to hot 
bottom burning.
\end{itemize}

Combining all evidences cumulated so far, it is clear that NGC~1851 has had a complex
history, as we also concluded in our previous analyses of the RGB (Carretta et al. 2010b 2011a) and
HB (Gratton et al. 2012b). The results obtained 
throughout this paper might be interpreted in a scheme where NGC~1851 is the result of the 
merging of two globular clusters, as suggested by Van den Bergh (1996), Catelan (1997), and 
Carretta et al. (2010c, 2011a). This might explain why we found separate O-Na anticorrelations
along the RHB and the BHB, and C-N anticorrelations along the b-SGB and f-SGB (see also
Lardo et al 2012). The apparent inconsistency that the older sequence is more metal-rich than the 
younger one can be understood if NGC~1851 formed within a chemical inhomogeneous structure 
and the two populations originated in different regions of this parent object. However, it is 
also possible that there is a precise relation between these two episodes of star formation.
The two most well-studied clusters exhibiting a spread in their subgiant branches, 
NGC~1851 and M~22, share a surprisingly long list of common features. In both cases, (i) the 
f-SGB appears more metal-rich than the b-SGB; (ii) they have similar [Ca/Fe] ratios; 
(iii) assuming a constant C+N+O, the age difference is about 0.5-0.6 Gyr; (iv) spreads in 
the abundances of C and N (at least) are present among both SGBs, with on average more 
N residing in the f-SGB stars (either owing to a larger fraction of N-rich stars or to a larger 
CNO content); and (v) n-capture elements are overabundant in the f-SGB relative to 
the b-SGB. It may well be that this long list of coincidences is not due only to chance, but they
are also consistent with both NGC~1851 and M~22 being simply the result of the merging of two 
globular clusters that formed independently.

\begin{center}
\begin{figure}
\includegraphics[width=8.8cm]{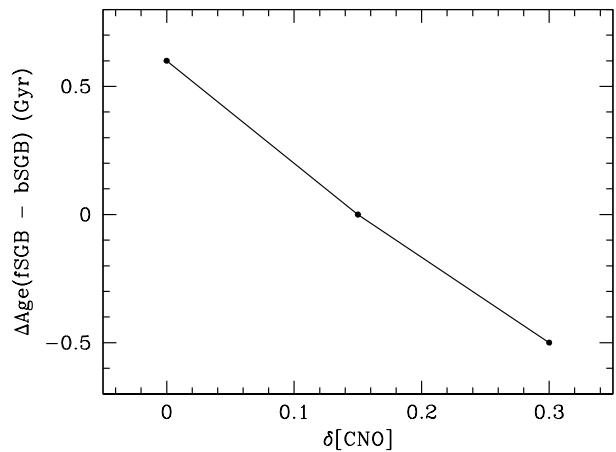}
\caption{Difference between the best-fit ages for the b-SGB and f-SGB of NGC~1851 as a 
function of the assumed difference in CNO abundance $\delta$CNO between the two populations.  }
\label{f:agecno}
\end{figure}
\end{center}

A crucial piece of information concerns the sum of C+N+O abundances. The difference
in the ratio (C+N+O)/Fe does not need to be huge to have an important impact on this
scenario: a mere difference of a factor of two in the sum of C+N+O abundances between 
b-SGB and f-SGB makes the latter actually younger than the first (see Figure~\ref{f:agecno}). 
In this case, we might 
possibly devise a scenario where a sequence of star formation episodes related to each 
other in a causal way generates all populations observed in both NGC~1851 and M~22. 
In this case, an important role is likely played by nucleosynthesis in stars with a mass 
of 2.5-4~$M_\odot$\ that might explain the abundance pattern of C-rich stars.
Unfortunately, when we need to sum the abundances of different elements, we must consider
absolute -not simply differential- abundances. Absolute abundances are much more sensitive to 
systematic errors, which can significantly affect the conclusion drawn. Perhaps 
unsurprisingly, the most recent results in the literature lack consensus. Various authors 
(Yong and Grundahl 2008; Yong 2011) found that there is a real difference in the ratio 
(C+N+O)/Fe, though their results can actually appear to be quite different depending on the 
observable that is used. On the other hand, no difference was found by Villanova et al. 
(2010), and a quite low upper limit was obtained by Gratton et al. (2012b).

The most easily observed stars are of course those on the RGB. Unfortunately, their
atmospheres are quite poorly understood, and the absolute abundances that have been
determined for these stars are likely
uncertain at a level possibly unacceptable in this context. In addition, the distinction between
the RGB progeny of b-SGB and f-SGB stars is not easy. A more robust determination of N 
abundances as well as a completely new determination of O abundances for dwarfs and
subgiants in NGC~1851 would clearly be highly welcome because it would enable us to
derive this badly needed datum. The use of CN bands, though observationally 
easier, is difficult, because if our errors in C abundances are underestimated, a spurious 
C-N anticorrelation can appear. It is unclear whether this is the case for the C-N 
anticorrelation found by Lardo et al. (2012), which might also be real. However, the use 
of an abundance indicator that is  completely independent of C abundances, such as the NH bands 
in the near UV is in our opinion preferable.

Finally, we propose that a detailed spectroscopic study of the HB of M~22 should be 
performed, in order to understand whether there is also bimodality as in the case of
NGC~1851, and to derive He abundances at least for some stars, as was done in M~4 by 
Villanova et al. (2012).

\begin{acknowledgements}
This research has made use of the NASA's Astrophysical Data System. This research 
has been funded by PRIN INAF "Formation and Early Evolution of Massive Star Clusters". 
S.V. and D.G. gratefully acknowledge support from the Chilean {\sl Centro de 
Astrof\'\i sica} FONDAP No. 15010003 and from the Chilean Centro de Excelencia en 
Astrof\'\i sica y Tecnolog\'\i as Afines (CATA). We thank A. Milone for having 
provided details about the ACS photometry. We also thank an anonymous referee
for useful suggestions that improved this paper.
\end{acknowledgements}

\end{document}